\documentclass[conference, letterpaper, 10pt, times]{IEEEtran}

\IEEEoverridecommandlockouts
\usepackage{cite}
\usepackage{adjustbox}
\usepackage{multirow}
\def\etal{\emph{et al.}}
\def\ie{\emph{i.e.,}}
\def\eg{\emph{e.g.,}}
\usepackage[ruled,vlined,linesnumbered,noresetcount,titlenumbered]{algorithm2e}
\usepackage{balance}
\usepackage{tcolorbox}
\usepackage[hidelinks]{hyperref}

\usepackage{amsmath,amssymb,amsfonts}
\usepackage{balance}

\begin{document}
\date{}
\title{\Large \bf On-Device Voice Authentication with Paralinguistic Privacy}
\author{
    \IEEEauthorblockN{Ranya Aloufi , Hamed Haddadi, David Boyle}
    \IEEEauthorblockA{Imperial College London}
    
}

\maketitle
\thispagestyle{empty}
\begin{abstract}
Using our voices to access, and interact with, online services raises concerns about the trade-offs between convenience, privacy, and security. The conflict between maintaining privacy and ensuring input authenticity has often been hindered by the need to share raw data, which contains all the paralinguistic information required to infer a variety of sensitive characteristics. Users of voice assistants put their trust in service providers; however, this trust is potentially misplaced considering the emergence of first-party `honest-but-curious' or `semi-honest' threats. A further security risk is presented by imposters gaining access to systems by pretending to be the user leveraging replay or `deepfake' attacks. 

Our objective is to design and develop a new voice input-based system that offers the following specifications: local authentication to reduce the need for sharing raw voice data, local privacy preservation based on user preferences, allowing more flexibility in integrating such a system given target applications privacy constraints, and achieving good performance in these targeted applications. The key idea is to locally derive token-based credentials based on unique-identifying attributes obtained from the user’s voice and offer selective sensitive information filtering before transmitting raw data. Our system consists of (i) `VoiceID', boosted with a liveness detection technology to thwart replay attacks; (ii) a flexible privacy filter that allows users to select the level of privacy protection they prefer for their data.
The system yields 98.68\% accuracy in verifying legitimate users with cross-validation and runs in tens of milliseconds on a CPU and single-core ARM processor without specialized hardware. Our system demonstrates the feasibility of filtering raw voice input closer to users, in accordance with their privacy preferences, while maintaining their authenticity.
\end{abstract}

\section{Introduction}
Voice Agents (VAs) offer a human-like means of interacting with users by understanding and producing speech.
~Today, there are a number of commercially available and deployed voice agents spanning services like banking, call centers, and medical services, in addition to smart assistants like Amazon Alexa and Apple's Siri. These agents often rely on verifying the user through speaker recognition (voice activation/spoken keyword) and then using speech recognition and natural language processing techniques to understand the spoken commands~\cite{hoy2018alexa}. 

Voice agents detect their `wake' word locally (i.e., on device) and automatically capture and send voice data,~\ie~commands to the cloud where they are interpreted and acted upon. The voice signal is an information rich resource that may disclose sensitive user attributes, such as emotional state, confidence and stress levels, physical condition, age, gender, and personal traits~\cite{singh2019profiling}, and may be used to profile users and thus further compromise privacy. Privacy concerns involve the use of personal and sensitive voice information, as classified by the GDPR, for purposes other than those intended. Amazon, for instance, has patented technology that can analyze users' voices to determine emotions and/or mental health conditions~\cite{Amazon_patent, 2204.10920}, neither of which is required to deliver the primary service(s). Moreover, leaking these voice personal attributes pose a risk of voice cloning and further security threats~\cite{yan2022survey, 9733178, seymour2018your, sukaitis2021building, wenger2021hello, papernot2016transferability, gomez2022adversarial}.

Anonymization techniques~\cite{tomashenko2022voiceprivacy, aloufi2021paralinguistic, 8844600, aloufi2020privacy, ahmed2020preech} can lower the probability of successful re-identification/reconstruction attacks, however, it comes with the cost of forgoing data authenticity. In this paper, we show that current privacy-preserving proposals have potential shortfalls when applied to interactive/real-time voice agents with cloud-based backends (Section~\ref{sec:2.3}). Thus, our objective is to design a system that offers~\emph{local security} (e.g., defeats impersonation or replay attacks),~\emph{cloud privacy} (e.g., minimizes curious inference or reconstruction attacks), and~\emph{reasonable privacy-utility trade-off} (i.e., maximizes the accuracy of the task of interest).


To meet our objective, we design and implement a system at the source that performs local user authentication using voice input that also offers configurable privacy preservation. To authenticate users, we first capture and store an on-device fingerprint of their raw signals rather than sending/outsourcing raw voice data. We then analyze this fingerprint using different predictors, each evaluating input authenticity from a different perspective (\eg~target identity claim, impersonating attempt, and liveness). We fuse multiple predictors' decisions to make a final decision on whether the user input is legitimate or not. Based on fused decision, we use this fingerprint to generate a private-public key which serves as an authenticity code. By doing so, service providers are able to verify their users without gaining access to raw user data that may compromise their privacy. Simultaneously, our solution can offer an additional layer of anonymity by applying an anonymization technique over the raw speech data before offloading such data to the cloud backends (Section~\ref{sec:3.3}). With our design, (1) service providers will not be required to authenticate their users using raw voice, (2) users will have sovereignty over their own data and keep their interaction private and secure~\cite{garrido2022revealing}, and (3) voice agents can be flexibly tailored to each application context (\eg~based on authenticity and privacy protection constraints per application). Our solution allows users to remain biometrically anonymous while their authenticity can still be verified by service providers.
\newline
\newline
\newline
\newline
\newline
\textbf{Contribution.} Our contributions can be summarized as follows:\\
\emph{\textbf{$\blacksquare$ Trustworthiness of Voice Agents.}}
Our systematic analysis (Section~\ref{sec:2.3}) shows that current voice agents fail to simultaneously achieve sufficient security and privacy. In particular, we examine whether proposed privacy-preserving solutions are compatible with existing voice agents without compromising their functionality. 
Our analysis uses current voice transformation tools applied to group of 20 participants and shows they can be used to imitate up to 97.5\% 
of speakers. Testing with agents including~\emph{Amazon Alexa},~\emph{Google Assistant}, and~\emph{Apple’s Siri} shows that these tools put the integrity of the agent and the security of users at risk. 
\newline
\emph{\textbf{$\blacksquare$ New Composable Design.}}
We design and implement a \emph{local} voice authentication system with selective privacy preservation, establishing trusted communication between service providers and their users. We present a new `{VoiceID}' that combines multiple predictors' decisions to ensure the authenticity of the user. Our new fusion score for user authentication achieves 98.68\% accuracy in validating legitimate users without compromising their privacy. Flexible privacy output retains its utility with minimal performance penalties of approximately 6\% word error rate (WER) compared to current cloud-based systems (Section~\ref{sec:3.3}). Following local authentication, a unique public-private key pair is generated on a user's device and the public-key is shared with the service provider. 
\newline
\emph{\textbf{$\blacksquare$ End-to-end evaluation \& real-world deployment.}}
We empirically evaluate the proposed system and systematically analyze its performance on edge devices, including a MacBook Pro i7 and a Raspberry Pi~4 boards. We demonstrate that the proposed system can effectively perform low-latency authentication and offer fair privacy protection on representative devices in tens of milliseconds (Section~\ref{sec:experiment}).\footnote{Code and research artefacts will be open sourced on acceptance for publication.}


\section{Automated Voice Agents}
\label{secrion:automated_voice_agents}
Users interaction with voice agents begins with some kind of trigger/activation such as `Okay, Google', `Alexa', and `Hey, Siri' to inform the system that speech-based data will be received. Voice activation might also include users authentication to restrict access to various systems to legitimate users. There are two phases involved in this process: enrollment and recognition. Users submit their biometric (voice) representations as part of the enrollment process to the service provider who stores them together with the user's ID in a central database. In the test phase, the user requesting access to a particular service will submit a new representation for authentication to the service provider. In response to the identity claim, the service provider retrieves the enrolled representations for comparison. Only if the two representations are close enough under a certain distance metric or threshold (\eg~cosine similarity~\cite{dehak2010front}), the user is successfully authenticated.

\begin{figure}
  \centering
  \includegraphics[width=0.49\textwidth]{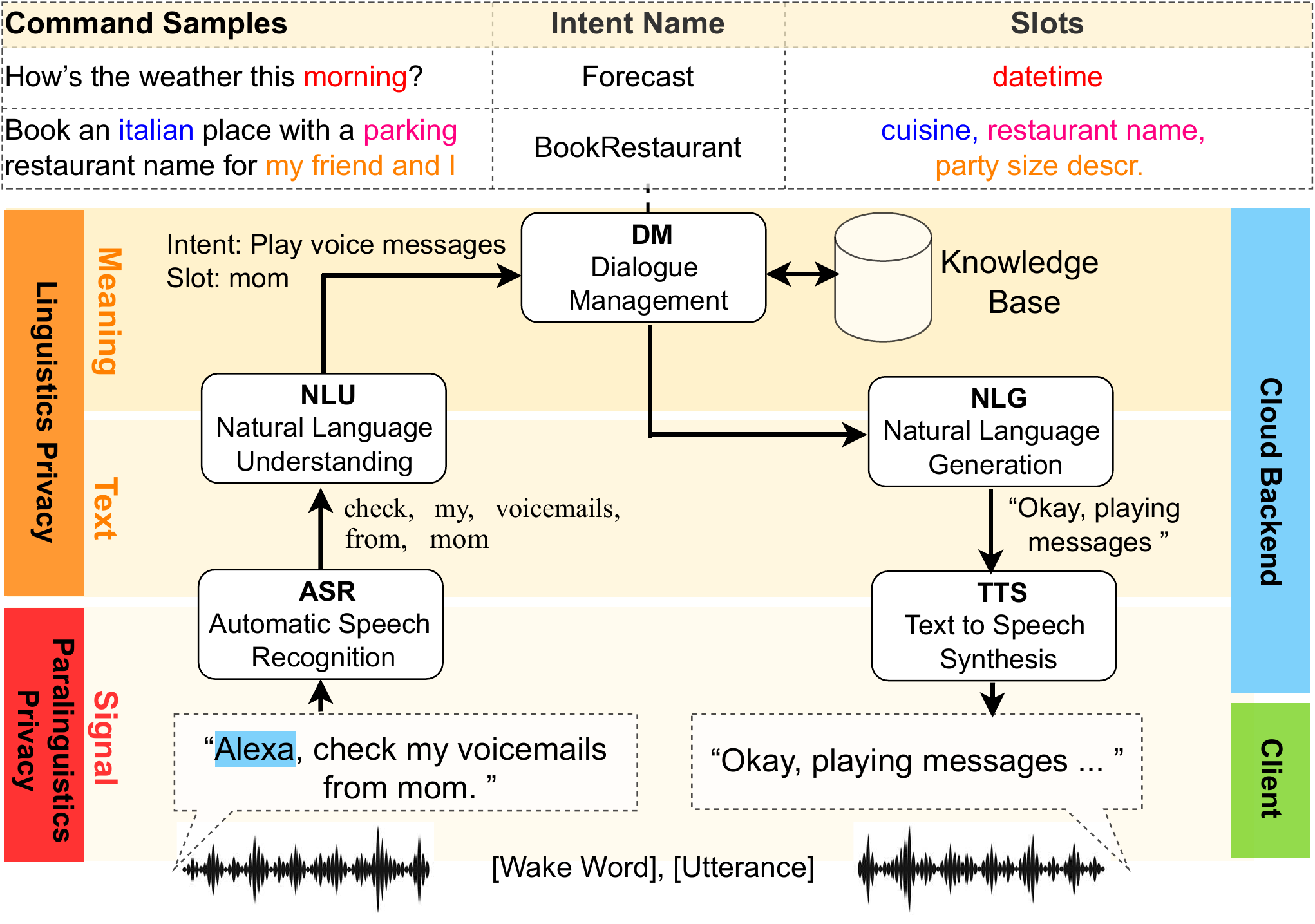}
  \caption{Typical user interaction with voice agents.}
      \label{fig:user_interaction}
      \vspace{-5mm}
\end{figure}

Once activated, these agents capture and outsource the raw voice data to more powerful cloud services such as automatic speech recognition (ASR) and natural language understanding (NLU) where higher performance is achievable. Hence, the user's voice information is used both for activating/authenticating and communicating with these agents. Figure~\ref{fig:user_interaction} shows an overview of how these systems work~\cite{perez2011conversational}. The ASR component converts acoustic user input into text, and passes the text string to the NLU component for semantic interpretation. Many approaches and models have been proposed for obtaining semantic meaning from speech~\cite{de2008spoken}. This involves typical NLP tasks such as named entity recognition, intent classification, and slot filling~\cite{burtsev2018deeppavlov}. The next step is for the dialogue manager (DM) to evaluate and/or disambiguate the semantic information obtained from the NLU module. This is achieved by examining dialogue history and interpreting contextual information. Following the evaluation of input, the DM plans and executes certain dialogue actions, such as making database queries or formulating system prompts. By using the natural language generation (NLG) module, the DM output (response) is transformed into a well-formed written utterance, which is then converted into voice by the text to speech (TTS) module.

\section{Problem Overview}
\subsection{Our Motivation}
The EU General Data Protection Regulation (GDPR) and the California Privacy Rights Act (CPRA) data minimization principle stipulate that only data necessary to fulfill a particular purpose can be collected~\cite{voigt2017eu, pardau2018california}. Since most of current voice-controlled services send raw data to cloud-deployed backends, collected raw data may be used to profile users and thus compromise their privacy further. Fundamentally, the problem lies in understanding how to introduce data outsourcing/sharing without allowing first parties to use sensitive or confidential data for non-designated uses. Users should also be able to choose which features they wish to share to enable a particular task or analysis on their data. Additional related issues that may arise are verifying the users’ authenticity without violating the individuals’ privacy as current systems can not do this. Typically, providers seek to protect their products and services, while users may seek to maintain the privacy of their personal information. Currently, it has been necessary to outsource raw data to enable authenticity, but this causes privacy concerns. 

\subsection{System \& Threat Models}
\textbf{\emph{System Model.}} Voice agent functions can be enabled natively by the host processor, or remotely using a client-server model. Currently, voice-controlled applications typically require client-server implementations. The client is responsible for detecting and encoding the voice input, packaging the data, and sending it to the server. The server processes audio data, decodes speech, and performs the necessary post-processing to improve the quality of the audio data. Users interact with these services through a connected microphone (which could be part of a wearable, earable, smartphone or smart speaker) that can collect their audio data. State-of-the-art voice agents typically work in two phases: the activation phase and the speech recognition phase. Once activated, the voice agents either start speaker-dependent speech recognition or speaker-independent speech recognition to recognize the voice commands. The former only processes voice commands spoken by a specific authorized user while the latter accepts voice commands from any speaker.

\textbf{\emph{Threat Model.}} Spoken commands can broadly convey information at multiple levels~\cite{holt2010speech} which can be categorised as (1) a linguistic level that pertains to the meaningful units of information structure in the speech signal, including phonemes, words, phrases, and sentences~\cite{holt2010speech} and (2) a paralinguistic/extralinguistic level that refers to non-verbal phenomena, including speaker traits, emotion, sex, accent, and ethnicity~\cite{singh2019profiling}. The utterance of a single word such as `security', for example, conveys indexical characteristics such as gender, familiarity, emotion, and sociolinguistic background of the speaker. It also conveys information about the phonetic category `/sIkyUUHRItEE/' which we link it to our semantic knowledge of security~\cite{aslin2014phonetic}. 
Reveling paralinguistic level information enables obtaining deeper insight into the user's behavioral patterns, which can be exploited to serve highly targeted content~\cite{Amazon_patent}. Amazon's recent announcement is a good practical example of potential security and privacy threats by using paralinguistic attributes to mimic anyone's voice~\cite{7}, which in addition to security threats via reply attacks, may contribute to new privacy risks. A potential threat is that an attacker can duplicate users's unique `voiceprint', allowing it to unlock voice agents or issue malicious commands, gaining access to protected resources (\eg~online banking accounts such as HSBC, Barclays, and Santander, making a purchase, or controlling a connected appliances~\cite{hoy2018alexa}). We intend to thwart \emph{first} party `honest-but-curious' service providers' attempts to infer sensitive attributes from raw audio data transmitted to large cloud-based language models for processing during voice interactions~\cite{malekzadeh2021honest}. Our approach involves implementing `privacy at the source' for data minimization~\cite{goldsteen2022data} which entails filtering sensitive attributes during the sensing process before using them within the system or decoding them for cloud storage (e.g., using the example of separating linguistic content from speaker-specific attributes).
\begin{figure}
  \centering
  \includegraphics[width=0.9\columnwidth]{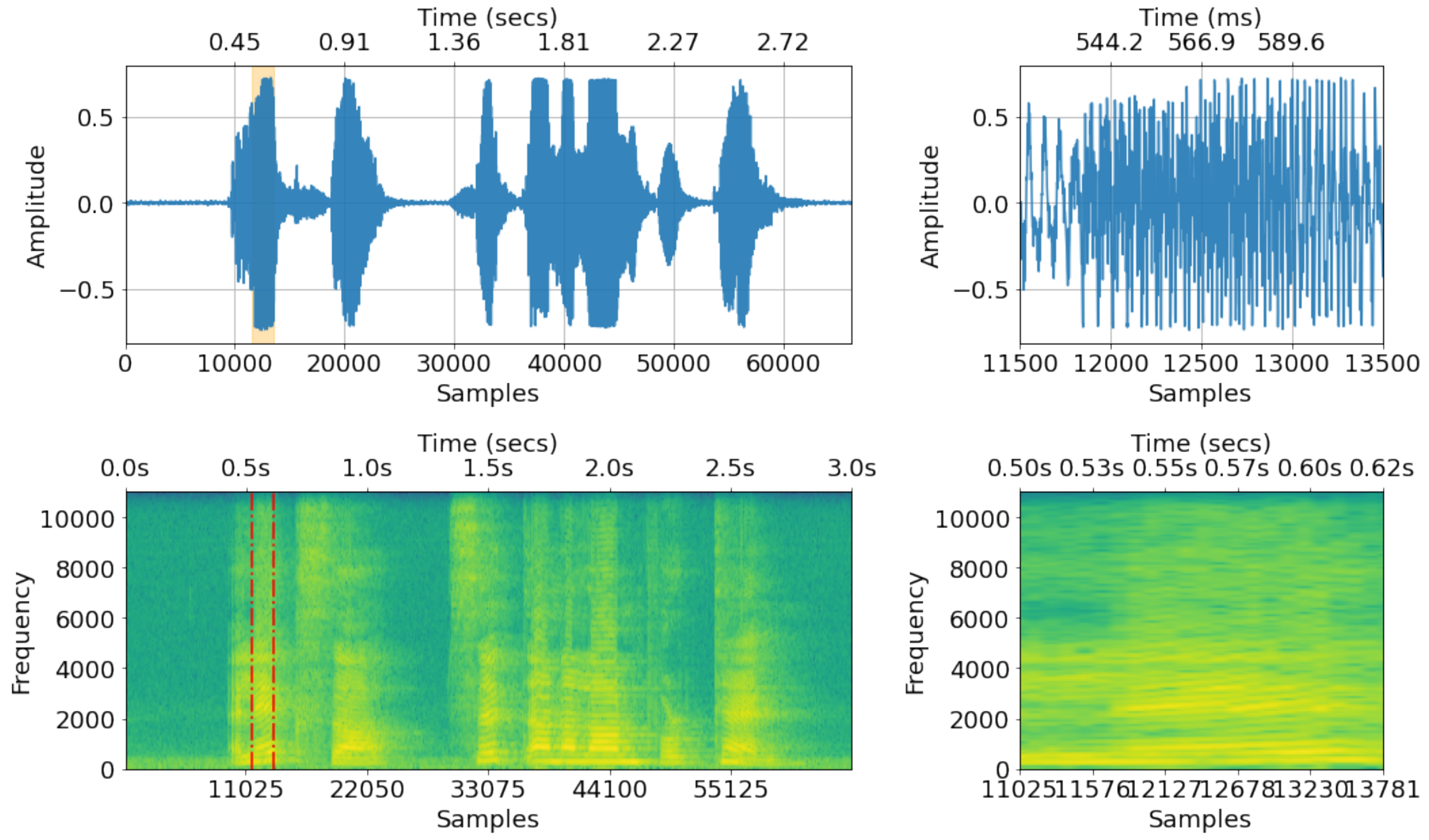}
  \caption{Real voice signal issuing the command ``Alexa, set an alarm for 10 pm." }
      \label{fig:signal_analysis_1}
      \vspace{-5mm}
\end{figure}
\begin{figure}
  \centering
  \includegraphics[width=0.9\columnwidth]{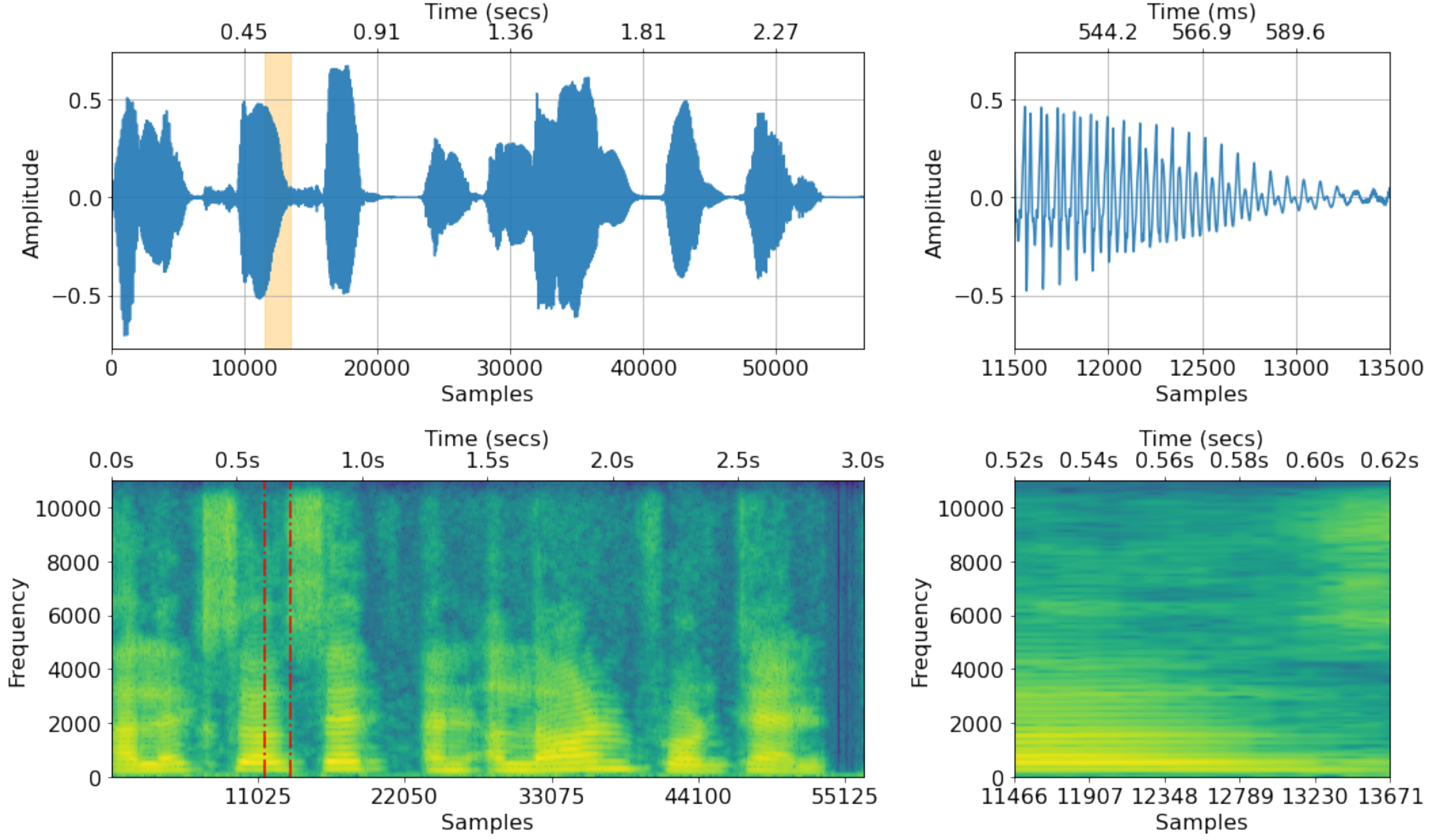}
  \caption{Fake voice signal issuing the command ``Alexa, set an alarm for 10 pm." }
      \label{fig:signal_analysis_2}
      \vspace{-5mm}
\end{figure}

\subsection{Our Technical Contribution}
To design trusted voice agents, we require: (1) robust voice authentication that verifies the authenticity of user input during the activation process while preventing impersonation attempts; (2) ensuring that sensitive behavioral and biometric characteristics in spoken commands are protected from untrustworthy service providers to assure users sovereignty; and (3) designing a resource-efficient solution that can operate effectively at the edge. To meet the above requirements, we present a local voice biometric-based authentication and flexible privacy preserving system as a holistic solution toward building trustworthy voice agents. We show that it is possible to deploy efficient privacy protection from the edge. We use and optimize existing state-of-the-art models/protocols to 1) reduce the amount of personal data needed to perform transcription tasks by filtering the input features of the runtime voice data and 2) show the feasibility that our `VoiceID' can be integrated with current voice agents without affecting its functionalities. Ultimately, we intend `VoiceID' to be deployed as close to the user as possible and give them control over what is shared. Achieving the right balance will depend on how user preferences affect the performance of interactions between these assistants and associated services (e.g., a privacy-utility trade-off).
\section{Deficiencies in Current Privacy Protection Solutions}
\label{sec:2.3}

We analyze current privacy-preserving voice analytics solutions in the context of interactive voice agent applications. Below, we demonstrate how they may fail to mitigate the potential privacy and security threats associated with real-time, seamless, and continuous interaction with users. 

\subsection{Privacy-preserving Voice Input}

\emph{\textbf{Current Solutions.}}
Voice transformation is one solution to mitigate privacy concerns, aiming to make the raw input unlinkable by altering a raw signal and mapping the identifiable personal characteristics of a given user to another identity (anonymity)~\cite{tomashenko2020introducing}. It usually consists of two stages: the first is to extract the speech features, suppress/anonymize speaker-related attributes, and then use a vocoder to convert it into speech~\cite{oord2016wavenet, tomashenko2022voiceprivacy, aloufi2019emotionless, aloufi2020privacy, sukaitis2021building, 9524669}. Such solutions usually lead to lower computational complexities due to their relatively simple operations compared to encryption-based mechanisms~\cite{zigomitros2020survey}. Encrypting the sensitive data using encryption schemes such as Homomorphic encryption aim to process encrypted information by untrustworthy parties without disclosing confidential information~\cite{9046801, 8884162, kaissis2020secure}. These encryption algorithms were excluded from our analysis due to its significant computational complexity, which makes it unsuitable for use with real-time voice agents. 

\textbf{\emph{Synthesizing Spoken Command.}}
Our analysis involved a total of 20 participants (Section~\ref{sec:experiment}). We consider low-resource settings given that we may only have access to fewer than 5 minutes of the target recordings, which might be not enough to train text to speech (TTS) models that need huge training data~\cite{wang2017tacotron}. Building high-quality TTS systems typically requires a large amount of high-quality paired text and speech data~\cite{tan2021survey}. To generate the spoken commands, we use open-source voice modeling tools `FastPitch'~\cite{9413889} and `HiFi-GAN'~\cite{NEURIPS2020_c5d73680} to train user voice models (\ie~20 models in total) and generate more realistic and engaging voices to the listener. FastPitch is a fully-parallel text-to-speech model based on `FastSpeech'~\cite{9053086NEURIPS2019_f63f65b5}, a fast, robust, and controllable (i.e., generated voice speed and prosody) text-to-speech tool. `HiFi-GAN' consists of one generator and two discriminators: multi-scale and multi-period discriminators. The generator and discriminators are trained adversarially, along with two additional losses, achieving generality to unseen speakers~\cite{NEURIPS2020_c5d73680}. We train two types of synthesizing models to meet the requirements of the two scenarios: one by using our participants' identities to generate target-specific samples (Section~\ref{sec:deceiving}) and another using fake/random identities to generate privacy-transformed samples (Section~\ref{sec:disabling}).

\subsection{Deceiving VUIs}
\label{sec:deceiving}
\textbf{\emph{Assumption.}}
We assume that deploying 
voice transformation tools (e.g., voice conversion) without any constraints can be exploited for malicious purposes and may facilitate the spread of impersonation attempts. 
Voice replay attacks can be used to impersonate a user's voice and grant access to an attacker. This attack is different to adversarial attacks that add imperceptible perturbations to the input sample to result in the incorrect prediction of the target system~\cite{9383529, carlini2018audio, abdullah2019practical, abdullah2019hear}. Such adversarial attacks are beyond the scope of our analysis.

\textbf{\emph{Setup.}}
Our analysis setup evaluates three voice agents:~\emph{Amazon Alexa},~\emph{Google Assistant}, and~\emph{Apple’s Siri}. The `VoiceID' of these agents links to to the primary user account, and thus we test the spoofed attack against these systems after setting up them to recognize our participants' voices. To create a `Voice Profile', a user repeats a list of Amazon-specified commands, and the profile is then linked to the primary user account. Google Assistant implements `Voice Match' to recognize who is speaking and deliver personalized results. Only one `Voice Match' profile can be associated with a Google account. To create a `Voice Match', a user says a few different phrases and their voice is processed to create a sonic fingerprint. Apple's Siri can be taught to recognize a user's voice and then uses it to serve up personalized content. 
Thus, Siri can recognize the target voice and tailor its responses accordingly. Once a device detects the keyword (\eg~`Alexa', `Ok, Google', and `Hey Siri'), it verifies that our participants could successfully use their real voices to log into and access these services. We use the trained models to generate 6000 fully-spoofed voice samples by feeding in target commands as text inputs. We also generate 2880 partially-spoofed commands~\cite{wagner2019speech}. In particular, we mix the spoofed `wake-word' utterance using our participants' voices with a random voice saying the rest of commands. We use state-of-the-art TTS framework `Coqui'~\cite{TTS}, a library for advanced Text-to-Speech models to generate the random voices. 

We then use an inexpensive JBL portable speaker located 0.5~m away from the devices to play participants' spoofed voices, with the process has repeated for each participant separately. We replay each command in Table~\ref{tab:spoofing_attack} once and record the responses by a target service. We use the attack success rate to evaluate how effectively spoofed voices can fool these agents. Since the commercial models verification systems are effectively a black-box, it is only feasible to assess the physical response on target or malicious activation attempts. An attacker succeeds if the target commercial service responds to spoofed voices the same way it responds to a real version of the commands.\\ 
\begin{table}[]
\small
\caption{Samples of command phrases (of cardinality 70) used in our experiments and the corresponding attack success (\%),~\ie~activating the service or not for two attacks settings: fully-spoofed (full command by target identity) and partially-spoofed (wake-word by target identity while the rest of command by random speaker).}
\label{tab:spoofing_attack}
\begin{adjustbox}{width=0.48\textwidth}
\begin{tabular}{|c|l|c|}
\hline
System & \multicolumn{1}{c|}{Commands} & \begin{tabular}[c]{@{}c@{}}Attack Success (\%)  \\  Fully / Partially \end{tabular} \\ \hline
\multirow{5}{*}{Alexa}  & Alexa, what's the weather today?           & 95 / 97.5 \\ \cline{2-3} 
                        & Alexa, set an alarm for 10 pm              & 85 / 87.5 \\ \cline{2-3} 
                        & Alexa, play song                           & 90 / 92.5 \\ \cline{2-3} 
                        & Alexa, what's in the news?                 & 80 / 87.5 \\ \cline{2-3} 
                        & Alexa, what's on my shopping list?         & 85 / 92.5 \\ \hline
\multirow{5}{*}{Google} & Hey Google, what's the time?               & 90 / 92.5 \\ \cline{2-3} 
                        & Hey Google, set a timer for 10 minutes.    & 80 / 87.5 \\ \cline{2-3} 
                        & Hey Google, what does my day look like?    & 70 / 87.5 \\ \cline{2-3} 
                        & Hey Google, what's the weather like today? & 75 / 87.5 \\ \cline{2-3} 
                        & Hey Google, call my phone.                 & 85 / 90.0   \\ \hline
\multirow{2}{*}{Siri}   & Hey Siri, find coffee near me.             & 90 / 92.5 \\ \cline{2-3} 
                        & Hey Siri, where's my iPhone?               & 85 / 90.0   \\ \hline
\end{tabular}
\end{adjustbox}
\vspace{-5mm}
\end{table}
\textbf{\emph{Evaluation.}}
Based on approximately two minutes of data, we demonstrate that a successful low-resource attack can be trained. Fraudulent `deepfaked' voices are shown to be sufficient to be granted access to and control over these commercial agents. On average, our spoofed attacks had 70-95\% success across all tests on these agents, as reported in Table~\ref{tab:spoofing_attack}. All 20 participants had at least 1 spoofed/faked command that fooled the tested services (\ie~\emph{Amazon Alexa},~\emph{Google Assistant}, and~\emph{Apple’s Siri}). These replayed/faked commands were able to access private shopping list and check calendar appointments. We found that a partially-spoofed voice (\ie~the victim's voice used only for the activation phrase) can give access to systems purportedly protected by voice profiles. This is a future system vulnerability, as such partial spoofed data might be streamed to take advantage of cloud services connected to these agents.

\subsection{Disabling Authentication Functionalities}
\label{sec:disabling}
\textbf{\emph{Assumption.}} We assume that when privacy-enhancing tools are applied to voice input, the resulting data is necessarily synthetic and no longer reflect the unique or true biometric characteristics of the individual that could be used to verify their authenticity. A question is thus raised regarding the feasibility of using such outputs to provide access to the voice agents' `VoiceIDs'. 

\textbf{\emph{Setup.}}
We select two state-of-the-art speaker verification systems: X-vectors~\cite{snyder2018x} and ECAPA-TDNN~\cite{desplanques2020ecapa} to evaluate the impacts on the authentication functionality. We use the raw recordings of our dataset as an enrollment set of the verification system. We then use the transformed voices as the test set. For the verification test, we compute the speaker embeddings from the enrollment and test sets and choose the threshold that minimizes the equal error rate (EER) for our target speakers, using cosine similarity as the distance metric. Authentication is considered successful if the similarity between the test and enrolled embeddings is above the threshold. For each user, we repeat the enrollment process 36 times (using different phrases samples) and report the average authentication success rate.

\begin{figure}[]
 \centering
   \begin{minipage}{0.5\columnwidth}
    \centering
     \includegraphics[width=\linewidth]{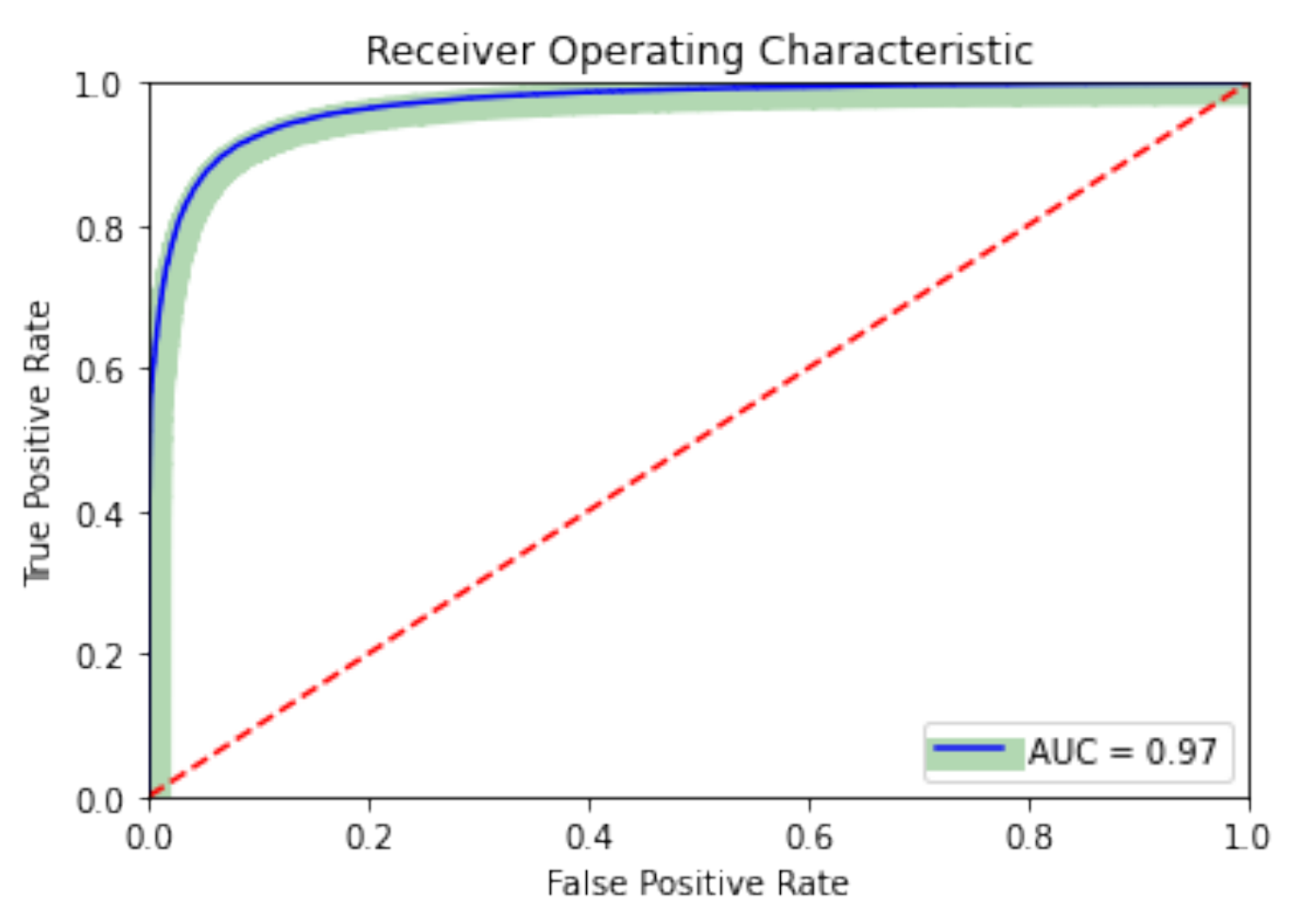}
   \end{minipage}
   \begin{minipage}{0.5\columnwidth}
     \centering
     \includegraphics[width=\linewidth]{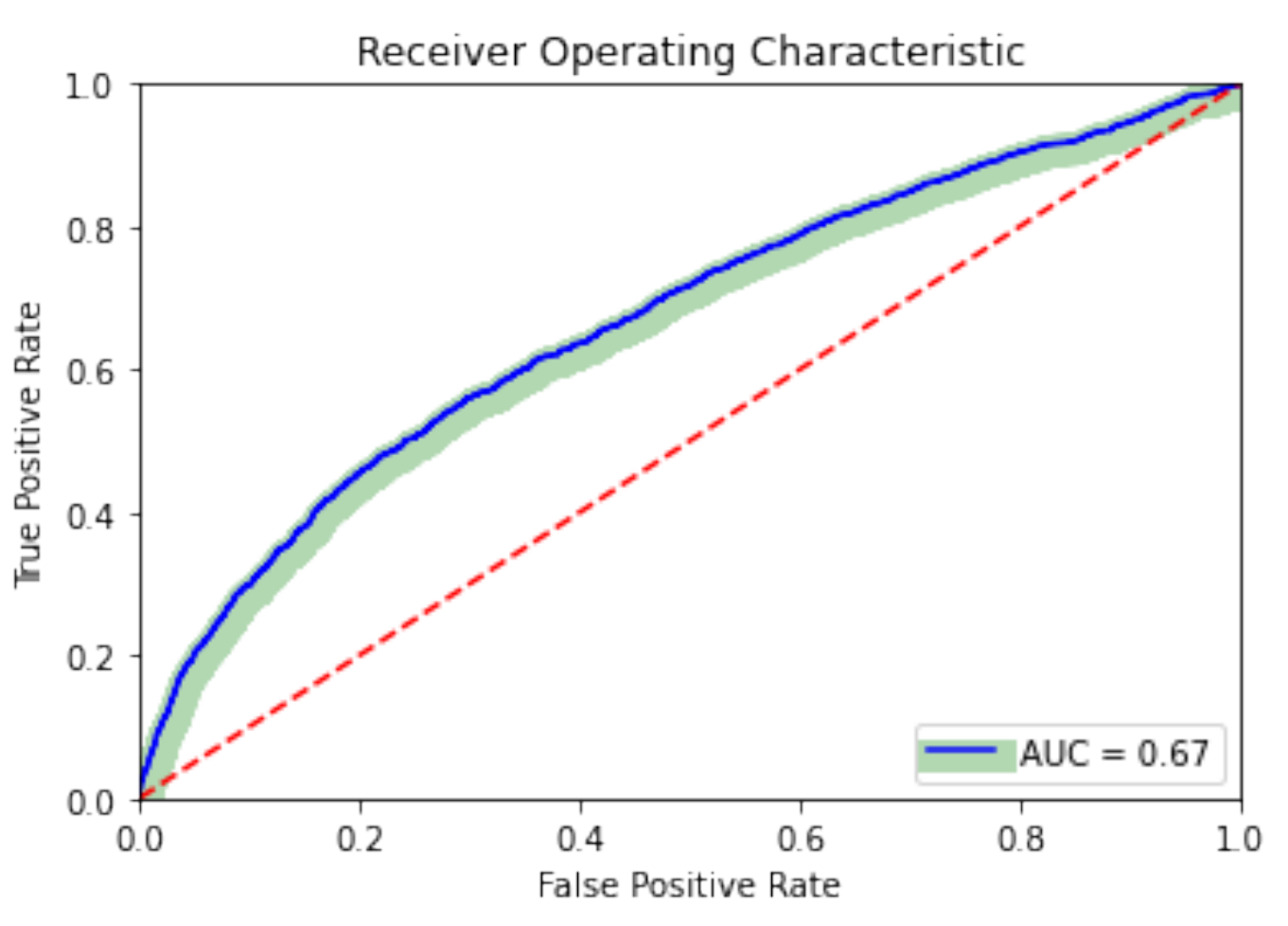}
   \end{minipage}
   \caption{Area under the curve (AUC) of raw samples (left) and privacy-transformed samples (right).}
   \label{Fig:AUC_eer}
   \vspace{-5mm}
\end{figure}

\textbf{\emph{Evaluation.}}
We tested a total of 6000 transformed voices targeting 20 speakers to test the speaker verification systems. We see from these results (Figure~\ref{Fig:AUC_eer}) that the use of voice transformation techniques showed good results in concealing the identity, aiming to achieving privacy by anonymity. However, this triggers questions about achieving privacy using synthetic data and whether such solutions investigate potential detrimental security consequences on voice agents, raising concerns about their applicability in real-time settings. This may open doors for new attacks related to trust between the service providers and their users.
\newline
\newline
\textbf{Takeaways.}
Currently, there are no voice agents that can provide both user authentication and privacy protection simultaneously on the user side~\cite{hoy2018alexa, yan2022survey, 9733178}. As we have shown above for current systems, malicious usage of the service will raise alarms concerning deceiving voice agents and gaining access to restricted resources, whereas benign usage will prevent service providers from verifying the authenticity of their users by blocking authentication functionality. Therefore, it is important that the authenticity of the user be incorporated into the privacy preservation pipeline, and to ensure that this verification does not interfere with the user's privacy. 

\section{Verifiable, Configurable Privacy Voice Input} 
\label{sec:framework}
\begin{figure*}
  \centering
  \includegraphics[width=\textwidth]{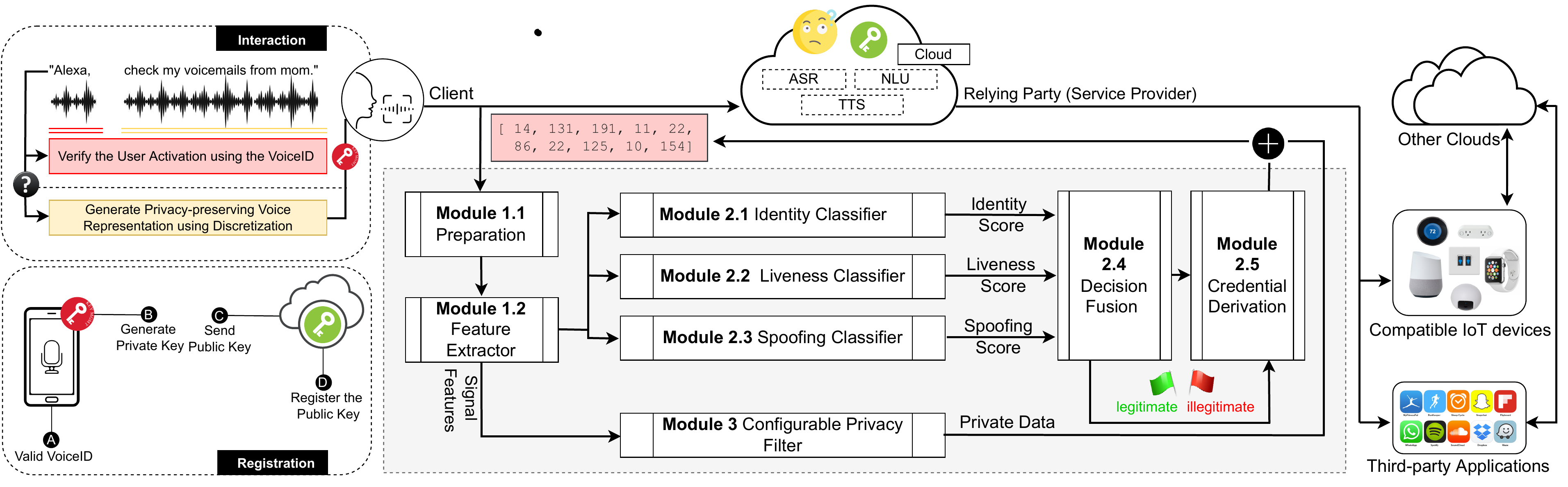}
  \caption{Client-side (i.e., on-device) architecture of our Local Voice Authentication with Selective Privacy System.}
      \label{fig:system_overview}
      \vspace{-3mm}
\end{figure*}

`VoiceID' is designed to address the conflicting needs of data owners and service providers: privacy processing without compromising security. Figure~\ref{fig:system_overview} illustrates an end-to-end overview of how users' spoken commands are processed using our solution. Voice input is used during the registration process to validate the user's authenticity (using the `VoiceID' module) and to generate public-private keys, storing the private key on the user's device and sending the public key to the service provider. During users interaction, their voice inputs passes through authentication and configurable privacy modules. Accordingly, the authenticity code is integrated into the privacy-preserving output (following steps 1.1 to 2.5). For example, if the user requests `Alexa, check my voicemail from mom', the wakeword will be used to invoke the private key. Depending on the user's privacy preferences, this command will also generate privacy-preserving voice input. Afterwards, the privacy-preserving output will be signed with the user's private key and forwarded to the cloud backend services for further analysis. Authenticity verification and privacy protection should be performed independently and simultaneously, thus increasing flexibility in designing voice agents based on the context of an application (for example, depending on its authentication and privacy constraints). Our design can eliminate the need for service providers to authenticate users using their raw voice biometric data. This can benefit users by giving them ownership/control of their own data and keep their information private and secure. We now describe the design elements of our ‘VoiceID’.

\subsection{Front-end Processing (Stage 1)}
\label{sec:3.1}
User interaction begins with capturing and pre-processing the user's voice data, i.e., commands. The feature extractor then extracts representative features, which can either be used to reflect user unique characteristics or serve as an input for privacy-preserving filter(s). Voice signal representations are extracted as follows:\\
$\blacksquare$ Acoustic features. Mel frequency cepstrum coefficients (MFCCs) represent the short-term power spectrum of a sound by linearly cosine transforming a log power spectrum on a nonlinear mel scale of frequency~\cite{zheng2001comparison}.  
The MFCC feature extraction technique includes windowing the signal, applying the DFT, taking the log of the magnitude, and then warping the frequencies on a Mel scale, followed by applying the inverse discrete cosine transform (DCT).\\
$\blacksquare$ Identity features. To extract the unique representations related to the user identity, we use a `deep speaker' model (\ie~using deep residual CNN (ResCNN) architecture) to extract frame-level features from utterances~\cite{li2017deep}. Then, affine and length normalization layers map the temporally-pooled features to a speaker embedding. The model is trained using triplet loss~\cite{schroff2015facenet}, which minimizes the distance between embedding pairs from the same speaker and maximizes the distance between pairs from different speakers.\\
$\blacksquare$~Inconsistency features. Artefacts that differentiate spoofs/replays from benign inputs can reside in the spectral or temporal domains. We first compute VOID features (\ie~97 features) for a given voice signal which includes the following four types of features: (1) low frequencies power features, (2) signal power linearity degree features, (3) higher power frequencies features, and (4) linear prediction cepstrum coefficients (LPCC)~\cite{mccandless1974algorithm} features, as its computational complexity is lower than MFCC because it does not require the computation of discrete fourier transforms~\cite{ahmed2020void}. Then, we use AASIST-L~\cite{Jung2021AASIST} which models both spectral and temporal information concurrently using a heterogeneous stacking graph attention layer to accumulate heterogeneous information.

\subsection{Authenticity Verification (Stage 2)}
`VoiceID' comes with pre-integrated classifiers to ensure that the voice input is verified from multiple perspectives towards robust authentication. To access and/or activate a voice agents, a legitimate input, which must be that of a `target user' and must be produced by a `live human' is required. These classifiers use traditional machine learning models to perform the training and testing on the previously extracted features (Section~\ref{sec:3.1}). Any feature-based classifier (\eg~logistic regression, decision tree, kNN, SVM, or neural network) may potentially be used. The input features of each classifiers is described in Section~\ref{sec:3.1} and the output of a classifier is the predicted score for each voice input sample. 
\newline
\textbf{Module 2.1: Identity Classifier.}
Verifying user identity involves comparing two inputs, namely the enrollment and testing inputs as $X$ = ($X_{enroll}$, $X_{test}$), where $X_{enroll}$ denotes a set associated with a known target identity and $X_{test}$ represents a single test sample. The output score (\ie~log-likelihood ratio) is denoted by $S_{id}$, and the threshold (operating point) is denoted by $T_{id}$. The final decision is then made upon the comparison of $S_{id}$ to a identity-specific threshold $T_{id}$: if $S_{id}$ $>$ $T_{id}$ then the target hypothesis is accepted. Otherwise, the non-target hypothesis is accepted. Given the voice command $X_{test}$, we use a pre-trained ‘deep speaker’ model with ResCNN architecture to extract embeddings that can be used as `identity features'. Then we compute the cosine similarity score $S_{id}$ between $X_{enroll}$, $X_{test}$ pairs where $X_{enroll}$ refers to the registered voice samples of target user and $X_{test}$ refers to voice command during user interaction.
\newline
\textbf{Module 2.2: Spoofing Classifier.}
Anti-spoofing works as a verification system by comparing a pair of inputs, namely the enrollment and testing inputs as $X$ = ($X_{enroll}$, $X_{test}$), where $X_{enroll}$ denotes set of samples corresponding to either genuine or spoofed speech and $X_{test}$ represents a single test sample. The spoofing countermeasure output score is denoted by $S_{spoof}$ and the threshold (operating point) is denoted by $T_{spoof}$. The spoofing decision is then made upon the comparison of $S_{spoof}$ to a spoofing-specific threshold $T_{spoof}$: if $S_{spoof}$ $>$ $T_{spoof}$ then the genuine hypothesis is accepted. Otherwise, the spoofed hypothesis is accepted. The spoofing detection model is text-independent and detects spoofed/deepfake attacks without requiring users to enroll specific passphrases. We compute the `inconsistency features (AASIST-L)' from the voice command $X_{test}$ using a pretrained AASIST-L model and use the model confidence score as spoofing indicator score $S_{spoof}$.
\newline
\textbf{Module 2.3: Liveness Classifier.}
Liveness measurement aims to detect a unique characteristics of the test input $X_{test}$ produced by an actual person and determine whether an input is from a live human or replayed. Thus, the liveness predictor is to reject all test signals that do not show evidence of liveness regardless of the nature of the replay attacks (\eg~speech synthesis or replay). The Liveness measurement output score is denoted by $S_{live}$. The liveness decision is then made upon the comparison of $S_{live}$ to a liveness-specific threshold $T_{live}$: if $S_{live}$ $>$ $T_{live}$ then the liveness hypothesis is accepted. Otherwise, the replayed hypothesis is accepted. Our liveness model is text-independent that verifies live users and detects replay attacks without requiring users to enroll specific passphrases. We first compute the `inconsistency features (VOID)' from the generated replayed samples of our dataset and then train a SVM classifier on top of these features. Given the voice command $X_{test}$, we use the pretrained model and output the classifier confidence score as $S_{live}$.
\newline
\textbf{Module 2.4: Decision fusion.}
Different from multi-biometric fusion methods~\cite{dinca2017fall} that use multiple biometrics for authentication, we combine the scores of different prediction tasks using the same modality. We consider it as a multi-view learning task~\cite{sun2013survey}. We aim at combining the confidence scores of the models constructed from different features, in which each confidence score measures the possibility of classifying a test sample $X_{test}$ into the positive class by one specific model. This is known as fusion at the measurement level or confidence level. Given a confidence score vector $s = [s_1,s_2,...,s_m]$ of a predictor model, where each $s_i$ denotes the score of the $i$th test sample, and $m$ is the sample number. The only classifiers discussed here are binary ones. All classifiers are assumed to return real values. A normalization step is required to adjust the weighting of each predictor to a common scale such that the combination can be performed. We aggregate the results of applying a number of binary classifiers to input data by leveraging the knowledge captured by each specific binary predictor. This allows using independent and possibly specialized classification techniques for each task. The final output score is denoted by `VoiceID' and the final decision on whether the user input $X_{test}$ is legitimate or not will be made based on this score.
\begin{equation}
\begin{split}
VoiceID = f_{fused} (S_{identity}, S_{spoof}, S_{live})
\end{split}
\end{equation}
\newline
\textbf{Module 2.5: Credential Derivation.}
The uniqueness of identity features (b) has been proven over the literature to be well-suited for verifying user authenticity~\cite{irum2019speaker}. Therefore, we show the potential for using these features (highly privacy-sensitive) for authentication without revealing any user data to the authenticating services. The FIDO2~\cite{WebAuthn} standard defines how to use public-key cryptography (PKC) for authentication. During registration/enrollment, the client creates a dedicated key pair per service as credentials with the help of a local authenticator (`VoiceID'). To register their credentials with an account, users send their service-specific public key to the service providers (`relying party')~\cite{schwarz2022feido}. During the interaction, an activation request signed with the user’s private key sent to relying party and the relying party will validate the signature of this request against the users public key and gain their access. A private key serves essentially as a proof of authentication. By doing so, service providers are able to verify their users without gaining access to raw user data that may compromise their privacy. The user's private key will be used to sign the privacy-preserving output, and the service provider will verify the signature against the user's public key (Details in Section~\ref{sec:authentication}). 

\subsection{Flexible Privacy Preservation (Stage 3)}
\label{sec:3.3}
In this paper we focus on real-time interaction with voice agents. Thus, paralinguistic privacy could be associated with the ASR component of these agents, while linguistic privacy could be associated with the NLU component. We investigate two common applications in which voice agents are used: smart speakers and online banking. The privacy concerns in these two applications are primarily related to paralinguistic privacy. It may be necessary to relax the linguistic privacy restrictions in the above contexts due to the possibility of misinterpretation or worse, e.g., changing the meaning of the intended texts~\cite{adelani2020privacy}. The above privacy constraints may not be sufficient for other applications. We thus emphasize the importance of enabling different privacy configurations for optimizing the privacy-utility trade-off. Such configurations might vary according to the users' preferences as to what they want to share with the service provider in given contexts. 

\textbf{Linguistic Privacy.}
Masking sensitive phrases or replacing them with a `bleep' noise is a common technique for protecting speech content privacy~\cite{williams2021revisiting}. It is possible, for example, to sanitize the content of an audio recording containing a list of target/sensitive words using automatic speech recognition (ASR) to obtain forced alignments in terms of words and timestamps. Afterwards, these sensitive words are masked with white noise, silence, or bleeps, and a new audio file is generated (audio de-identification). Such approaches might be useful in strengthening the anonymity objective when storing/transcripting voice data~\cite{ahmed2020preech, qian2019speech}, however they are not sufficient for real-time interaction with voice agents. A specific question to consider is how much privacy is actually protected by such text transformations, and whether the resulting texts are still useful for the functionality of these agents. Although the scope of this paper does not extend to this area, it may be an interesting direction for future research.

\textbf{Paralinguistic Privacy.}
For anonymity purposes, our target is to learn a representation useful for transcription tasks, and such representation not contain additional information that would allow secondary/curious tasks. Discrete units (\eg~phonemes) highlight linguistically relevant representations of the speech signal in a highly compact format~\cite{nguyen2020zero} while being invariant to speaker-specific and background noise details. One motivation to apply discretization is that these discrete units can capture high-level semantic content from the speech signal, \eg~phoneme due to the discrete nature of phonetic units. In~\cite{aloufi2020privacy}, the authors analyzed the effectiveness of learning such representations in promoting privacy protection and used the example of separating linguistic content and speaker-identity attribute. Inspired by this method, we use CPC-kmean clustering~\cite{nguyen2020zero} to extract the phonetic content (\ie~speaker-invariant). 
Once we tune this discretization model, the output can be connected to different downstream purposes,~\eg~Automatic Speech Recognition (ASR)~\cite{kepuska2017comparing, zhou2016speech}, Natural Language Processing (NLP)~\cite{hoy2018alexa, digan2021can}, and/or Speech Synthesis (TTS)~\cite{wang2017tacotron}. A response will likely be sent back to the user similarly to current client-server interaction. Using our system, it is possible to support any computationally efficient anonymization mechanism without compromising the user's authenticity.

\section{Experimental Settings}
\label{sec:experiment}
\emph{\textbf{New Dataset.}}
We recruited a total of 20 participants to create a new dataset whereby we could ensure full understanding of the ground truth, `liveness', and control of labelling. Using the same recording settings, each participant repeated each command three times from a prepared list of realistic voice assistant commands (12 commands). All of the voice samples were recorded at a sampling frequency of 48~kHz. The voice commands were mixed in length (approximately ranging from 2 to 6 seconds) and command types (\eg~setting alarms, asking for news and weather, and calling contacts). 55.14\% of the participants were male and 44.86\% were female, ensuring that both male and female voice frequency ranges were covered. The participants were in the 18-24 (33.33\%), 25-34 (47.62\%), and 35-44 (19.05\%) age groups. Our participants have different linguistic backgrounds (5 native English speakers (US/UK); 4 native Mandarin speakers; 3 native Middle-eastern speakers; 2 native Marathi speakers; 1 native Italian speaker; 1 native Russian speaker; 1 native Dutch speaker; 1 native Niger-Congo speaker; 1 native German speaker; 1 native Portuguese speaker). We explicitly informed the participants that the purpose of the voice sample collection was to develop and evaluate a secure `VoiceID' solution, with all institutional policies followed.

\emph{\textbf{System Prototype.}}
We implement the I/O module using librosa libarary v0.9 that includes read and write methods to read from the microphone input (signal collection) and write it in the form of a .wav file. In our experiments, we use a common sampling frequency of 16 kHz. Following common practice~\cite{neekhara2019expediting}, inputs are set to be log-mel spectrogram patches of 80 bins from the collected signals using librosa v0.9, with an FFT size of 2048 and hop size 256 as Mel-spectrogram parameters. We use state-of-the-art models to generate lightweight, optimized version suitable for edge devices. For our `VoiceID', we used three validation modules: deep speaker~\cite{li2017deep} for user identity verification, AASIST-L~\cite{Jung2021AASIST} for spoofed signal detection (logical), and VOID~\cite{ahmed2020void} for liveness checking whether the signal is live or recording (physical). Each standalone module's detection performance was measured using an equal error rate (EER), which are approximately 3.2\% (deep speaker model), 0.64\% (AASIST-L model), and 7.62\% (VOID model). For privacy-preserving module, we use CPC-kmean clustering~\cite{nguyen2020zero} to extract the phonetic content (\ie~speaker-invariant). To facilitate the deployment of the selected models on embedded devices, quantization~\cite{jacob2018quantization} has been applied to obtain a small enough model to run on-device. Our design is quite flexible and configurable, allowing developers to use any model of their choice, and integrate it without affecting the overall system structure.

\emph{\textbf{Hardware.}}
We conduct our experiments using a Z8 G4 workstation with Intel (R) Xeon (R) Gold 6148 (2.8 GHz) CPU and 256 GB RAM. The operating system is Ubuntu 18.04. We train and fine-tune all models on an NVIDIA Quadro RTX 5000 GPU. Then, we deploy the trained models on a MacBook Pro with an Quad-Core Intel i7 CPU and a Raspberry Pi 4B with a Broadcom BCM2711 CPU, quad core Cortex-A72 (ARM v8) 64-bit to simulate the specifications of current voice-controlled devices.

\emph{\textbf{Ethics.}}
All our study protocols were carefully designed in alignment with institutional regulations designed to protect the privacy and ensure the well-being of our participants. We retain only audio recordings that have been anonymized and stored on secure servers.

\section{Evaluation}
We apply the following evaluation criteria to assess the efficacy of our proposed solution:\\
\emph{$\blacksquare$ Security.} Verify the authenticity of the target user and prevent malicious/faked voice inputs.\\
\emph{$\blacksquare$ Privacy.} Measure the level of privacy protection offered to a user and the ability of an attacker to reconstruct sensitive attributes from the shared data.\\
\emph{$\blacksquare$ Utility.} Test the applicability of the proposed solution in real-world settings and its effectiveness.\\
\emph{$\blacksquare$ Efficiency.} Identify the computational overhead and resources required for the propsed solution to operate effectively from the edge.

\subsection{Authentication}
\label{sec:authentication}
\textbf{Setup.} 
Voice input from all sources should be subject to authenticity verification. The ideal case for an input to be categorized as authentic is to be verified as the target user, benign/real (\ie~not faked), and a live entry. Otherwise, input is considered non-authentic. Non-authentic inputs may occur if the user is not the target but may be live voice, and the worst case is that it is not of the non-target user, produced by machine and played from a recording. Our focus is to defend against attacks that input signals `sounding' like a target's voice to humans and machines alike. We first pre-processed the input (voice) and then extract the acoustic features and embeddings using the proposed solution's front-end. The features and embeddings extraction might be varied based on the tasks, for example, we used deep speaker for speaker embedding, and AASIST-L and VOID for spoofing and liveness embeddings respectively. These modules then calculate its own decision scores on given voice input. We extracted these scores for each sample (\ie~speaker similarity score, spoofing score, and liveness score) in our dataset. Each validation module (classifiers) can have its own multidimensional feature vector derived from the raw input. We produced approximately 117,000 scores combinations, simulating various input scenarios. After normalizing the scores, we trained a binary classifier to classify the input as authentic or not. During the fusion process, match scores outputted by different classifiers are consolidated in order to arrive at a final authentication decision. 

\textbf{Credential Derivation.}
We select a unique set of attributes from voice signals that represent the authenticity of the users, namely user identity features (b) (Section~\ref{sec:3.1}). We generate the credential by feeding these attributes into the key derivation function to derive user-specific credentials. We utilize a SHA-256 HMAC~\cite{krawczyk2010cryptographic} as a key derivation function to hash the attributes and then calculate the WebAuthn respective public key using the hash. This hash can be used as the private key of the user's WebAuthn credentials. It is possible to use this reproducible public key pair with the typical FIDO2 credential registration/authentication process~\cite{schwarz2022feido}. 
\begin{table}[]
\caption{Cross-validation results across models: mean values and standard deviations (in parentheses).}
\label{tab:score_fusion}
\begin{adjustbox}{width=0.5\textwidth}
\begin{tabular}{|c|c|c|c|c|}
\hline
Model & Accuracy        & Precision       & Recall          & F-score         \\ \hline
SVM   & 0.9863 (0.0005) & 0.9858 (0.0035) & 0.9863 (0.0052) & 0.9860 (0.0039) \\ \hline
MLP   & 0.9862 (0.0004) & 0.9850 (0.0129) & 0.9862 (0.0233) & 0.9858 (0.0087) \\ \hline
kNN   & 0.8980 (0.0125) & 0.9721 (0.0126) & 0.8980 (0.0158) & 0.9245 (0.0197) \\ \hline
SGD   & 0.9840 (0.0063) & 0.9832 (0.0458) & 0.9840 (0.0660) & 0.9825 (0.0644) \\ \hline
LR    & 0.9862 (0.0005) & 0.9856 (0.0039) & 0.9862 (0.0051) & 0.9858 (0.0040) \\ \hline
\end{tabular}
\end{adjustbox}
\vspace{-5mm}
\end{table}

\textbf{Evaluation.}
We apply the five classifiers (e.g., SVM, MLP, kNN, SGD, and logistic regression) as the final score predictor on our dataset to classify the user's legitimacy. We perform a k-fold cross-validation where the mean and one standard deviation of a 5-fold test for accuracy, precision, recall, and F score are presented in Table~\ref{tab:score_fusion}. All 5 classifiers achieve over 98\% Accuracy, with a Precision rate in excess of 97\%. The best model, SVM, achieves 98.63\% Accuracy, 98.58\% Precision, and 98.63\% Recall. The classification results have very small variances. This confirms that the fusion of multiple predictors are indeed useful for the user authenticity classification task. However, there is a possibility that the match scores generated by different predictors may not be homogeneous. For instance, one predictor may provide a distance or dissimilarity measure (a smaller distance indicates a better match), while another may provide a similarity value (a greater similarity shows a better match). This shows the importance of normalization during the fusion process to get a fair decision about the user authenticity. Upon this decision, a private key will be generated/unlocked, which is paired with a public key held by the service provider. There is no information available to the service providers regarding the method used to unlock the private key, only that it was used to sign the data. As a result, the users can verify their authenticity without having to share raw data.

\textbf{Security Analysis.}
By leveraging the unique attributes of users' identities extracted from their voices, we are able to generate pseudonymous authentication credentials in order to verify their authenticity. Schwarz~\etal~in~\cite{schwarz2022feido} provide a security analysis of this form of credentialing, as well as a discussion of its potential to strengthen system security without compromising user privacy. Similar to~\cite{schwarz2022feido}, we use HMAC in our implementation, which Bellare showed to be a pseudorandom function~\cite{bellare2015new}. Therefore, the security of propsed `VoiceID' follows from~\cite{schwarz2022feido, hanzlik2022token} to provide unlinkable FIDO credentials. 
Apple's Passkeys for Touch ID or Face ID are two practical examples of how public-private key authentication boosts online security~\cite{6}. In the case of voice agents, we are sending the raw data, whereas, for Touch ID and Face ID, we are only sending the generated security code. Accordingly, our system is designed to conform to the latter approach. Our experimental results show that we can achieve strong security ($>$98\% authentication accuracy) while balancing privacy preservation, utility and latency. There is clear room for improvement and further analyses, but we believe this offers a compelling entry point and benchmark for the community.

\subsection{Privacy Protection}
\begin{figure}
  \centering
  \includegraphics[scale=0.48]{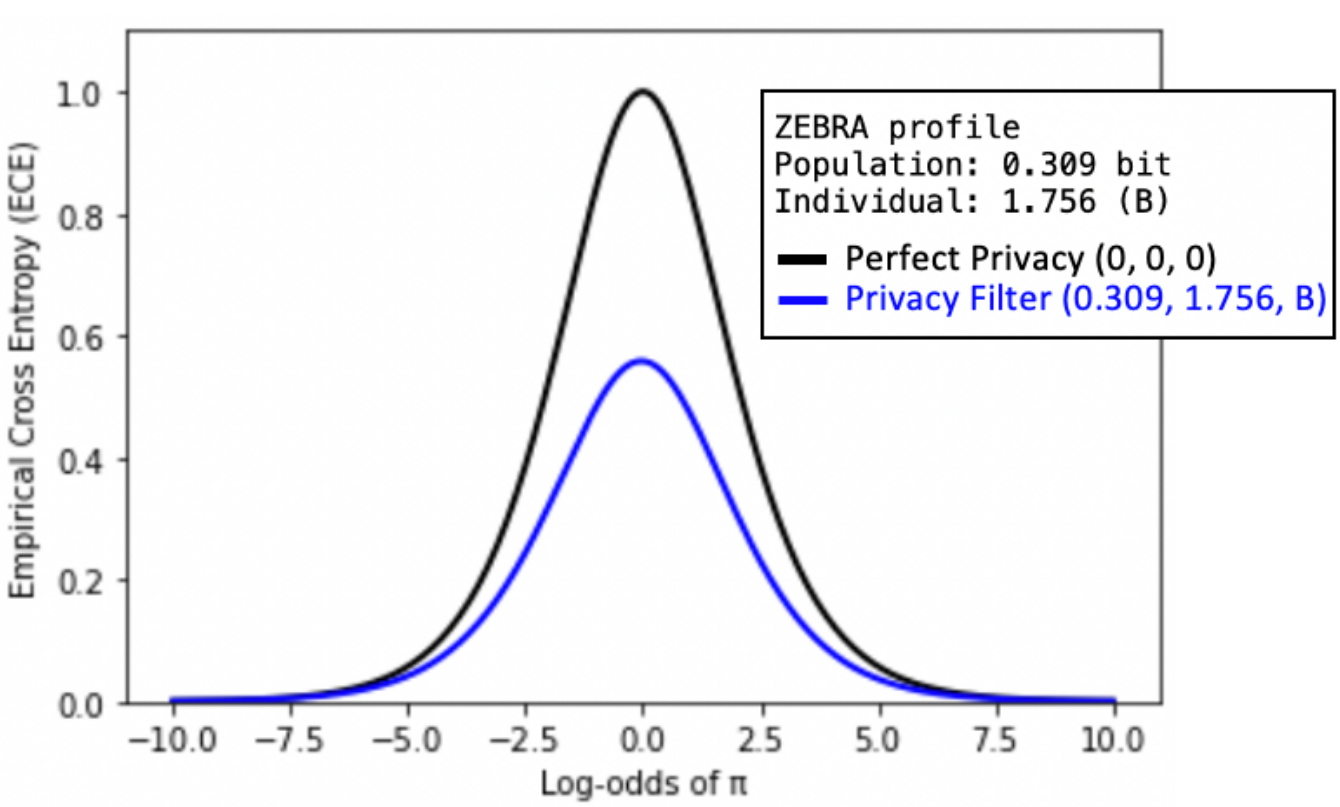}
  \caption{Anonymity level offered by the privacy preservation filter.}
  \label{fig:privacy_preservation}
  \vspace{-5mm}
\end{figure}

\begin{figure*}[htp]
\centering
\includegraphics[width=.30\textwidth]{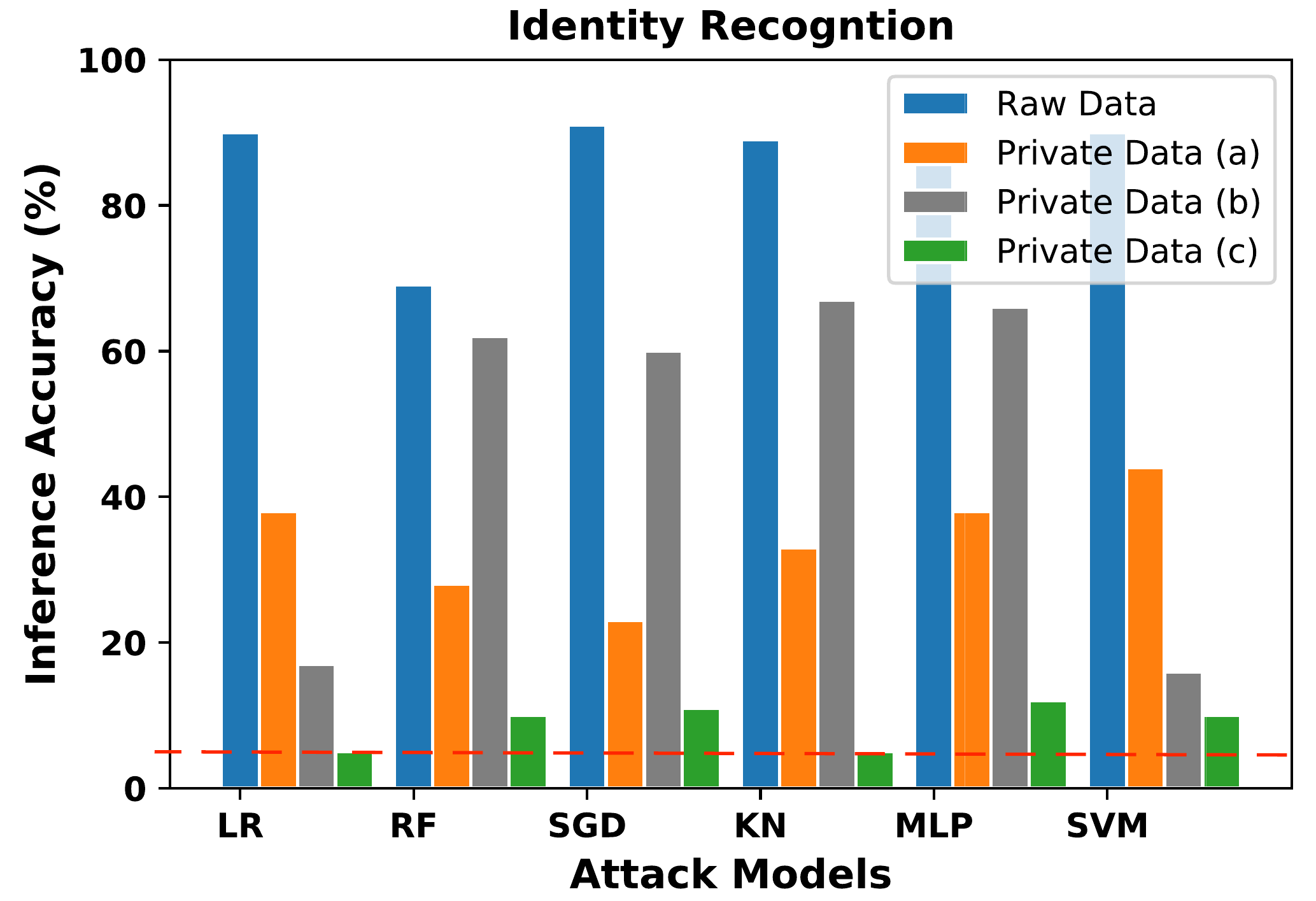}\quad
\includegraphics[width=.30\textwidth]{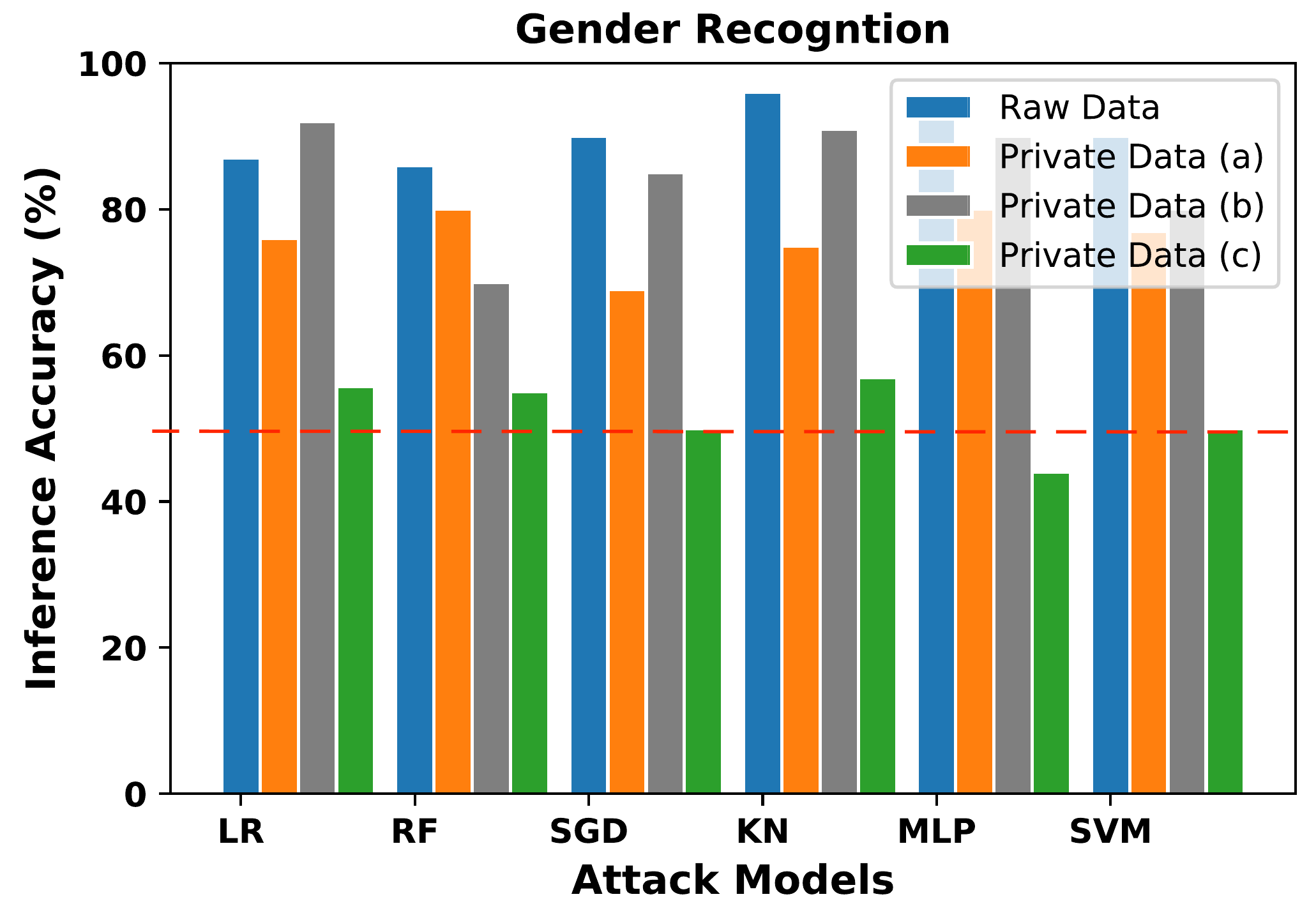}\quad
\includegraphics[width=.30\textwidth]{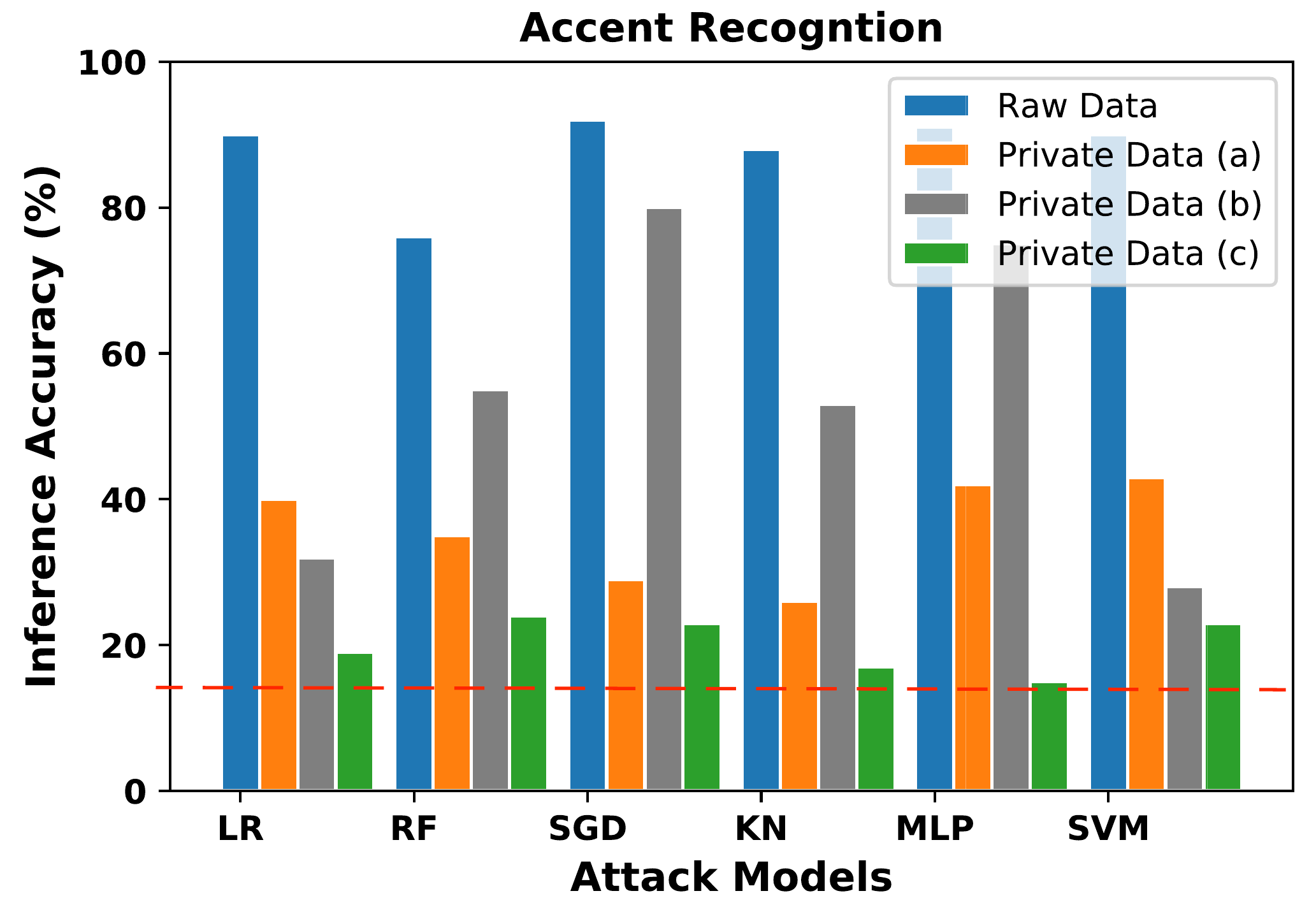}
\caption{Accuracy in inferring various attributes (\ie~identity, gender, accent) using both raw and private data (\ie~the output of the privacy-preserving tools ((a) signal-based anonymization~\cite{kai2021lightweight}), (b) voice privacy~\cite{tomashenko2020introducing}, and (c) disentanglement~\cite{aloufi2020privacy})). The red line represents a random guess for each attribute.}
\label{pics:privacy_analysis}
\vspace{-3mm}
\end{figure*}
\subsubsection{Paralinguistics Privacy}
We mainly focus on protecting the sensitive attributes that may be inferred from signals (e.g., identity, gender, and emotion) and are not required by the main functionality of real-time voice agents (understanding spoken commands). 
\newline
\textbf{$\blacksquare$ Inferring Sensitive Attributes.}
\newline
\emph{\textbf{Setup.}}
We consider an adversary with full access to user data with the aim to correctly infer sensitive attributes (\eg~identity, gender, and accent) about users. Aloufi~\etal~in~\cite{aloufi2020privacy} investigate the effectiveness of a similar attacker (\eg~a ‘curious’ service provider) who may use deep acoustic models trained for speech recognition or speaker verification to learn further sensitive attributes from user voice input. In our case, we assume that the privacy attack is an `honest-but-curious' service provider's effort to obtain additional information from the shared voice data by users that they did not intend or expect to share. To test the effectiveness of these attacks, we assume that the sensitive attributes in our dataset are identity, accent, and gender (\ie~available labels). An attacker trains a machine learning algorithm that takes the raw voice recordings as input, trying to infer the above attributes from these recordings. We test the success of such attack over binary (\ie~gender) and non-binary (\ie~identity and accent) attributes. For each of these attributes, we train spesific model using the raw recording of our dataset. This allows us to demonstrate the potential leakage of sensitive attributes caused by current practices in sharing raw data with the providers of voice agents. Then, we evaluate the success of such attacks over the privatized version of these recording using best-performing privacy-preserving voice analytic systems, \ie~(a) signal processing-based anonymization~\cite{kai2021lightweight}, (b) voice privacy baseline (TDNN-based)~\cite{tomashenko2020introducing}, and (c) disentanglement~\cite{aloufi2020privacy}. We retrain the attacker's classifiers and report their success comparing with non-filter (raw) recordings. We measure the success of these attacks by the increase in inference accuracy over random guessing~\cite{wagner2018technical}.

\emph{\textbf{Evaluation.}}
The success rate for a variety of attacks is presented in Figure~\ref{pics:privacy_analysis}. We show that inference models have varying performance, ranging from about 40.6\% to 80.2\% in successfully inferring different attributes from the raw data. The attacker has the opportunity to extract sensitive information with a much higher degree of accuracy than would otherwise be possible by chance. For example, for identity recognition, and assuming that we have 20 different speakers, then the random assumption rate will be $\sim$5\% of the time, but when using the `MLP' model the success rate is four times greater than this. We then measure an attack's success as the increase in inference accuracy over random guessing~\cite{yeom2018privacy}, and compare this with the inference success of the raw data as our baseline. Comparing to the inference success from raw data, the performance of the used privacy protection methods vary from one system to another. We found that our implementation for disentanglement by quantizing the voice input (Private Data (c), Figure~\ref{pics:privacy_analysis}) offers the best performance. It is approximately in line with guessing at random for all attacker models. 
\newline
\textbf{$\blacksquare$ Anonymity Measurement.}
\newline
\emph{\textbf{Setup.}}
Our privacy-preserving filter is evaluated to determine whether it is able to maintain the biometric anonymity of shared voice data. In our case, `identity' is the sensitive property that we want to protect (\ie~part also of user's authentication). We applied the anonymization method to the raw recordings of our participants to hide their identities and then reproduced the signal. We use the speaker verification system decision score that indicates the similarity of the given utterance with target identity. Then, we use the output scores within the `ZEBRA' framework~\cite{Nautsch_2020}. ZEBRA measures the average level of anonymity afforded by a given privacy-preserving solution for a population and the worst-case privacy disclosure for an individual~\cite{Nautsch_2020}. These metrics can be interpreted using categorical tags, and the odds ratio for recognizing a biometric identity by the lowest $l$ value of a category. Lower odds means less precision for an adversary, and thus more privacy preservation (best is 50:50 with a category `A').

\emph{\textbf{Evaluation.}}
For privacy protection level, the higher the percentage of dissimilarity between the raw data before and after anonymization, the better the identity protection we have. In Figure~\ref{fig:privacy_preservation}, the black curve corresponds to perfect privacy or zero evidence, where the blue line represents the result (0.309, 1.756, B) in the form of a ($D_{ECE}$, $log_{10} (l)$, tag) tuple, where the $D_{ECE}$ (\ie~empirical cross entropy (ECE)) provides an average protection estimation afforded to a population whereas $log_{10} (l)$ provides additional insights about the protection level afforded in a worst-case to an individual. Our results demonstrate that we can achieve a fair level of privacy (\ie~ label `B') from the edge using lightweight techniques.

\subsubsection{Linguistics Privacy}
\begin{figure}
  \centering
  \includegraphics[scale=0.46]{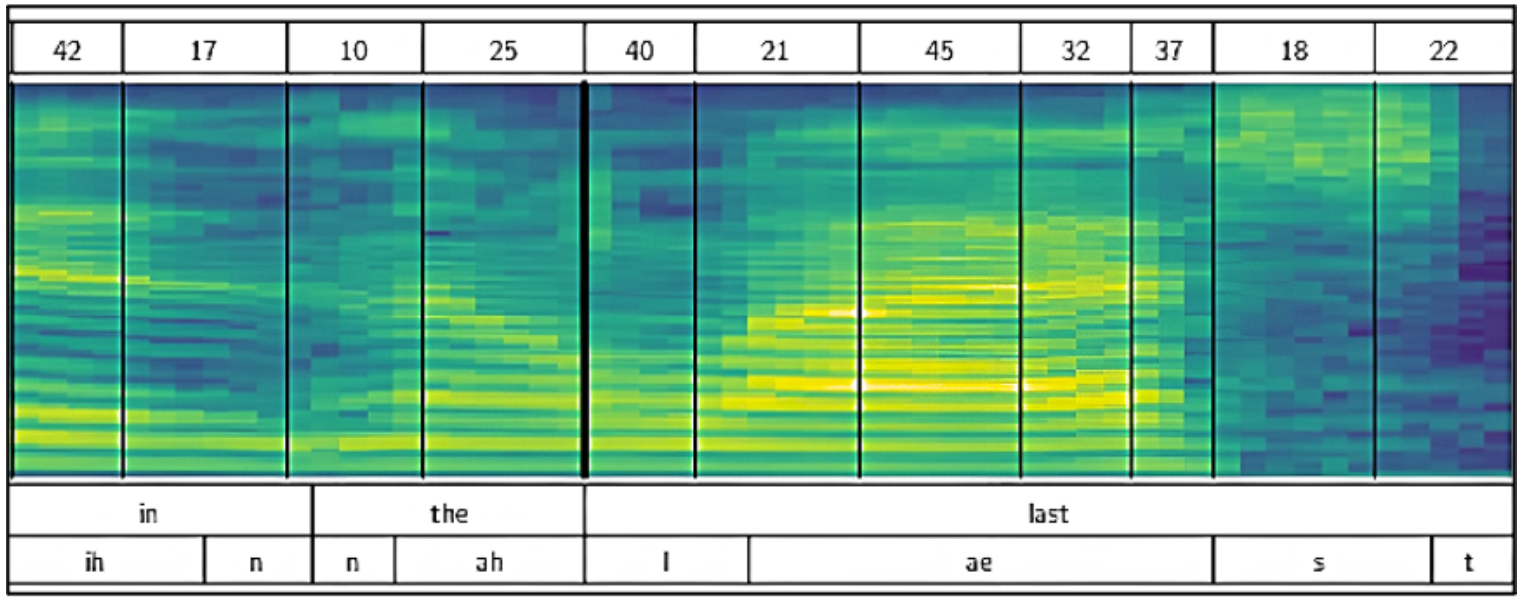}
  \caption{Word segmentation output shown on signal's spectrogram; top row indicates discrete units of the CPC-kmean, while bottom rows show words and phonemes ground-truth.}
      \label{fig:word_segmentation}
      \vspace{-5mm}
\end{figure}

Typically, intent prediction and slot filling are considered to be two NLU tasks that learn to model intent labels (sentence level) and slot labels (word level)~\cite{weld2021survey}. Amazon's Alexa, Apple's Siri, and Google Assistant use such techniques extensively for goal-oriented dialogue systems. Effective conversational assistance depends on identifying the `intent' behind a query and identifying relevant `slots' in the sentence to engage in a dialogue.
For example, users might want to ‘play music’ given the slot labels ‘year’ and ‘artist’. We mentioned in Section~\ref{sec:3.3} that one method to achieve linguistic privacy is to filter out the sensitive words (slots) contained in the input voice data, while the rest of the stream remains unaltered (audio de-identification~\cite{cohn2019audio}). For the sake of demonstration, we use word segmentation to find fragment onsets and offsets that align with word boundaries so that further privacy filters can be applied. In particular, a symbolic word segmentation algorithm is used to produce a discrete low-bitrate speech signal segmentation that uses the output of discretization models as input, inspired by~\cite{kamper2020towards}. This algorithm can detect boundaries without any constraints on segment length or the number of segments in an utterance. Figure~\ref{fig:word_segmentation} is an example. The top of this figure shows the code indices from CPC-kmean clustering (dicrete units)~\cite{nguyen2020zero} plotted on the input spectrogram. We believe that the next step toward linguistic privacy is the implementation of policies to allow the de-identification of sensitive data in real time based on the users privacy preferences and application contexts. 
However, this is beyond the scope of this paper.

\subsection{Utility Evaluation}

\textbf{Setup.}
Utility is measured by the ability to maintain the main functionality of the online service when using the privacy-preserving inputs. In our example, in voice-controlled interfaces, understanding and responding to voice commands (\ie~automatic speech recognition (ASR)) is the primary task of these services. We use ASR-based metrics to evaluate the quality of the filtered data to demonstrate the proposed framework's feasibility and compatibility with transcription systems. We use state-of-the-art ASR systems to translate the generated speech back to text and then apply metrics including word error rate (WER) to determine the intelligibility of the resulting speech in terms of higher linguistic content. This reflects that we still have a privacy-preserving version of the raw audio that is sufficiently good for the transcription task. To demonstrate the practical applicability of the proposed framework with current cloud-based models (commercial Speech-to-Text APIs), we use a subset of the Librispeech test dataset (raw recordings) as a baseline, assuming that such recordings disclose all sensitive information about the user. We measure the WER, which is the ratio of edit distance between words in a reference transcript and the words in the output of the speech-to-text engine to the number of words in the reference transcript (\ie~lower WER means the more precise is the model), and the real-time factor (RTF), which is the ratio of CPU (processing) time to the length of the input speech file. A speech-to-text engine with lower RTF is more computationally efficient. We use the ground-truth transcripts within the dataset to calculate the WER of the raw (baseline) and our framework output (privacy-aware generation).

\textbf{Evaluation.}
We calculate the word error rate (WER) using the Automatic Speech Recognition (ASR) of current speech-to-text cloud-based services and uses the ground-truth transcripts. Examples include Amazon Transcribe~\cite{Amazon}, Google Speech~\cite{Google}, IBM Watson~\cite{IBM}, Mozilla DeepSpeech~\cite{Mozilla_DeepSpeech}, and a local transcription model trained on the Librispeech dataset. We can see from Table~\ref{tab:asr_performance} that utility is maintained with minimal performance penalties of~$\sim$6\% word error rate (WER) compared to current cloud-based ASR systems.

\subsection{Real-time Performance}
\textbf{Setup.} 
Processing and verifying inputs at the source (\eg~at the smartphone or smart speaker) offers a means to counter security and privacy attacks from their onset, and thus we examine~\emph{if we can apply input verification and filtration on resource-constrained devices.} In our experiment, we deployed both authenticity verification and flexible privacy module (i.e., disentanglement) on two representative edge platforms: a MacBook Pro and a Raspberry Pi 4B. We report the average inference times, number of features used, and the average memory required by these modules.

\textbf{Evaluation.} 
As shown in Table~\ref{tab:performance}, the results indicate that we can deploy these models on the different edge/cloud devices with promising overall inference time and memory usage in all cases. The results demonstrate the feasibility of applying our system in the real world. It is shown to be efficient for input authenticity checking, where the total time required is no more than 700 ms in our experiments. The inference time roughly increases linearly with the length of the tested recording. Memory consumption on the MacBook Pro and Raspberry Pi 4 is 0.006 kB and 0.002 kB, respectively. 
\begin{table}[]
\caption{The word error rate (WER; lower is better) and real-time factor (RTF; lower RTF is more computationally efficient).}
\label{tab:asr_performance}
\begin{adjustbox}{width=0.48\textwidth}
\begin{tabular}{|l|cc|cc|cc|cc|cc|}
\hline
\multicolumn{1}{|c|}{\multirow{2}{*}{Service}} &
  \multicolumn{2}{c|}{Amazon} &
  \multicolumn{2}{c|}{Google} &
  \multicolumn{2}{c|}{IBM} &
  \multicolumn{2}{c|}{Mozilla} &
  \multicolumn{2}{c|}{Local Model} \\ \cline{2-11} 
\multicolumn{1}{|c|}{} &
  \multicolumn{1}{c|}{WER} &
  RTF &
  \multicolumn{1}{c|}{WER} &
  RTF &
  \multicolumn{1}{c|}{WER} &
  RTF &
  \multicolumn{1}{c|}{WER} &
  RTF &
  \multicolumn{1}{c|}{WER} &
  RTF \\ \hline
Raw Data &
  \multicolumn{1}{c|}{17.26} &
  57.49 &
  \multicolumn{1}{c|}{18.45} &
  7.35 &
  \multicolumn{1}{c|}{7.14} &
  2.11 &
  \multicolumn{1}{c|}{5.95} &
  1.33 &
  \multicolumn{1}{c|}{5.36} &
  2.20 \\ \hline
Private Data &
  \multicolumn{1}{c|}{23.21} &
  51.41 &
  \multicolumn{1}{c|}{85.00} &
  2.87 &
  \multicolumn{1}{c|}{32.74} &
  1.39 &
  \multicolumn{1}{c|}{27.38} &
  1.04 &
  \multicolumn{1}{c|}{11.31} &
  2.06 \\ \hline
\end{tabular}
\end{adjustbox}
\vspace{-5mm}
\end{table}

\section{Related Work}
\textbf{Privacy-preserving Speaker Verification.}
The problem of protecting privacy in speaker verification (user authentication) has been a major research area in the speech community. 
One of the earliest approaches is cryptographic-based such as homomorphic encryption and secure multi-party computations~\cite{pathak2012privacy, portelo2014privacy, nautsch2019privacy, treiber2019privacy}. Other approaches have explored how to use distance-preserving hashing techniques for privacy-preserving speaker verification~\cite{Portelo2013secure, MTiBAA2021Privacy, Fan2019Deep}. Differently, Teixeira~\etal~in~\cite{teixeira2022towards} emphasize the need to achieve security while protecting both the client's data and the service provider's model. Most of the current secure speaker authentication schemes employ a server-centric model where a service provider maintains a voice biometric storge and is responsible for ensuring the security of the voice biometric representations. Users must therefore trust the server to store, process, and manage their private representations. However, the legitimate receiver might reverse engineer the received data (re-identification attack), making these techniques insufficient for ensuring privacy when the recipient is not fully trusted (`honest-but-curious')~\cite{garrido2022revealing, 10.1145/3460120.3484533}. In contrast to the above works, our work focuses on real-time interaction between users and voice agents that includes both activation and sharing functions. Our solution for filtering personal voice data locally without having to send complete raw data to backend language models over the internet, where it could be misused (i.e., a \emph{first party} threat to privacy), thereby minimizing data leakage risks over shared data.
\begin{table}[]
\small
\caption{The computational cost by different framework's modules including: speaker verification (SV), spoofing detection (SD, logical), liveness detection (LD, physical), and privacy-preserving (PP) on two edge representative platforms, namely MacPro (Pro) and Raspberry Pi4 (Pi).}
\label{tab:performance}
\begin{adjustbox}{width=0.48\textwidth}
\begin{tabular}{|c|l|clclclcl|}
\hline
\multicolumn{1}{|l|}{\multirow{3}{*}{}} &
  \multicolumn{1}{c|}{\multirow{3}{*}{Measure}} &
  \multicolumn{8}{c|}{Module} \\ \cline{3-10} 
\multicolumn{1}{|l|}{} &
  \multicolumn{1}{c|}{} &
  \multicolumn{2}{c|}{SV} &
  \multicolumn{2}{c|}{SD} &
  \multicolumn{2}{c|}{LA} &
  \multicolumn{2}{c|}{PP} \\ \cline{3-10} 
\multicolumn{1}{|l|}{} &
  \multicolumn{1}{c|}{} &
  \multicolumn{1}{c|}{Pro} &
  \multicolumn{1}{c|}{Pi} &
  \multicolumn{1}{c|}{Pro} &
  \multicolumn{1}{c|}{Pi} &
  \multicolumn{1}{c|}{Pro} &
  \multicolumn{1}{c|}{Pi} &
  \multicolumn{1}{c|}{Pro} &
  \multicolumn{1}{c|}{Pi} \\ \hline
\multirow{2}{*}{ {Time (s)}} &
  Extraction &
  \multicolumn{1}{c|}{0.34} &
  \multicolumn{1}{l|}{0.84} &
  \multicolumn{1}{c|}{0.53} &
  \multicolumn{1}{l|}{0.67} &
  \multicolumn{1}{c|}{0.12} &
  \multicolumn{1}{l|}{0.14} &
  \multicolumn{1}{c|}{0.04} &
   0.33 \\ \cline{2-10} 
 &
  Testing &
  \multicolumn{1}{c|}{0.10} &
  \multicolumn{1}{l|}{0.90} &
  \multicolumn{1}{c|}{0.01} &
  \multicolumn{1}{l|}{0.04} &
  \multicolumn{1}{c|}{0.58} &
  \multicolumn{1}{l|}{0.59} &
  \multicolumn{1}{c|}{0.05} &
   0.09\\ \hline
\multirow{3}{*}{{Memory (kB)}} &
  Features &
  \multicolumn{1}{c|}{192} &
  \multicolumn{1}{l|}{192} &
  \multicolumn{1}{c|}{128} &
  \multicolumn{1}{l|}{128} &
  \multicolumn{1}{c|}{97} &
  \multicolumn{1}{l|}{97} &
  \multicolumn{1}{c|}{100} &
   100 \\ \cline{2-10} 
 &
  Memory &
  \multicolumn{1}{c|}{2.97} &
  \multicolumn{1}{l|}{0.73} &
  \multicolumn{1}{c|}{0.03} &
  \multicolumn{1}{l|}{0.02} &
  \multicolumn{1}{c|}{3.50} &
  \multicolumn{1}{l|}{1.29} &
  \multicolumn{1}{c|}{0.09} &
   0.03\\ \cline{2-10} 
 &
  Model Size &
  \multicolumn{1}{c|}{0.86} &
  \multicolumn{1}{l|}{0.86} &
  \multicolumn{1}{c|}{25.3} &
  \multicolumn{1}{l|}{25.3} &
  \multicolumn{1}{c|}{2.6} &
  \multicolumn{1}{l|}{2.6} &
  \multicolumn{1}{c|}{55} &
   55 \\ \hline
\end{tabular}
\end{adjustbox}
\vspace{-5mm}
\end{table}
\newline
\textbf{Defeating Presentation Attacks.}
A presentation attack detection (PAD) system aims to determine the authenticity of the biometric presentation (voice)~\cite{sahidullah2019introduction}. 
In particular, it aims to detect any artifacts in the input that will match the nature of the deepfake, such as a noisy glitch, phase mismatch, reverberation, or loss of intelligibility~\cite{wang2021comparative}. Several features have been proposed to capture these artifacts~\cite{wu2012detecting, todisco2017constant, sahidullah2015comparison}.
Since the majority of these methods require access to the raw data in order to generate the verification score and decision, privacy concerns may arise. We implement them locally in our `VoiceID' to address such concerns.
\newline
\textbf{Biometric Passwordless Authentication.}
For passwordless solutions, FIDO2 (Fast Identity Online) is commonly used, which combines WebAuth and CTAP (Client to Authenticator Protocol). The FIDO2 protocol utilizes pairs of cryptographic keys instead of transmitting the authentication data. 
Currently, these passwordless solutions are used in conjunction with fingerprint scanning and face recognition, making them one of the most accurate authentication technologies available today~\cite{4, 5}. Consider that for voice-based systems we are sending the raw data, whereas in Face ID and Touch ID~\cite{5}, only the generated security code is sent. Our solution is designed to conform to the latter approach. Our experimental results show that we can achieve strong security (about 98\% authentication accuracy) while balancing privacy preservation, utility and latency. There is clear room for improvement and further analyses, but we believe this offers a compelling entry point and benchmark for the community.


\section{Conclusion and Future Research}
\textbf{Privacy Formalization for Voice Input.}
The highly sensitive nature of voice data~\cite{singh2019profiling}, which may contain unique identifiers beyond those necessary to perform primary tasks, is causing us to rethink the extent to which it should be shared and with whom. Privacy-preserving applications should only send task-based representations to the cloud to enable the main functionalities. Identifying the required level of privacy protection depends on the context of the application. In designing our system, we consider privacy to be subjective, where attitudes may differ among users depending upon the services (and/or providers) with which their systems communicate. A key part of improving the privacy-utility trade-off is enabling different privacy settings and promoting transparency of privacy management (flexible privacy). The theoretical comparison of privacy and utility is left to future research, since we need a framework/metric that unifies efforts to protect voice privacy, clarifies privacy restrictions, and considers security implications.\\
\textbf{Continuous Authentication.} 
We have evaluated our system on a limited number of subjects and the system will need to be evaluated with larger numbers of varied participants to better understand and improve performance. Long-term study may consider the possibility of individuals' characteristics changing over time, e.g., voice changing due to illness. The idea of continuous authentication is to establish the user’s identity not just once at login time but also continuously while the person is using the system~\cite{eberz201928}. 
Continuous authentication can be achieved by regular credential regeneration.\\ 
\textbf{Robustness Measures.} In our experiments, we focused on detecting impersonating attacks (i.e., spoofing and replay) based on the assumption that every modification to the underlying data must be disclosed. In this paper, we specifically aim to detect any attempts to use target voice biometrics to create an artificial version for the purpose of obtaining unauthorized access to sensitive or protected resources~\cite{7194916, wang2020asvspoof}. An attacker, for instance, may use a recording device to record a user's voice commands and replay the recorded samples using a stand-alone speaker to complete the attack. More robustness analyses can be carried out to evaluate the pipeline's effectiveness against other attacks, such as adversarial spoofing~\cite{das2020attacker} and hidden attacks~\cite{abdullah2019practical}. In such cases, the system may incorporate additional modules like ensemble for keyword spotting (EKOS) with the authentication to sharpen its robustness~\cite{ahmed2022towards}.\\
\textbf{Optimization versus Performance.}
One of the primary reasons for taking an edge computing approach is to filter data locally before sending it to the cloud. Local filtering may be used to enhance the protection of users' privacy. We were able to obtain light models for on-device deployment, but as future work we will seek to advance its deployment on devices with even more limited resources, e.g., by using more optimization, including model quantization~\cite{jacob2018quantization} and knowledge distillation~\cite{10.1145/1150402.1150464, hinton2015distilling}, to obtain even faster and smaller models. Further exploration of these optimization approaches for more constrained devices is left for future research.

\bibliographystyle{acm}
\balance
\bibliography{sample}

\end{document}